%% file: arxiv.tex
\documentclass[10pt] {article}

\usepackage[usenames,dvipsnames,svgnames,x11names]{xcolor}
\definecolor{gray}{rgb}{0.4,0.4,0.4}
\definecolor{lightgray}{rgb}{0.9,0.9,0.9}
\definecolor{maroon}{rgb}{0.5,0,0}
\definecolor{darkgreen}{rgb}{0,0.6,0}
\usepackage{bbding}

\newcommand{\Note}[1]{\textcolor{red}{#1}} 
\newcommand{\claim}[1]{\vspace{-1.35ex}\subsubsection{\textnormal{\textit{#1}}}}

\newif\ifsectionbreaks
\sectionbreaksfalse

\usepackage[T1]{fontenc} 
\usepackage[normalem]{ulem} 

\usepackage[export]{adjustbox}

\usepackage{listings}

\lstdefinelanguage{XML}
{
basicstyle=\ttfamily\footnotesize,
  morestring=[b]",
  moredelim=[s][\bfseries\color{Maroon}]{<}{\ },
  moredelim=[s][\bfseries\color{Maroon}]{</}{>},
  moredelim=[l][\bfseries\color{Maroon}]{/>},
  moredelim=[l][\bfseries\color{Maroon}]{>},
  morecomment=[s]{<?}{?>},
  morecomment=[s]{<!--}{-->},
  commentstyle=\color{gray},
  stringstyle=\color{blue},
  identifierstyle=\color{red}
}
\usepackage{moreverb}

\usepackage[nounderscore]{syntax}

\graphicspath{{./figures/}}
\DeclareGraphicsExtensions{.pdf}
\usepackage[export]{adjustbox}

\usepackage[cmex10]{amsmath}
\usepackage{amssymb}

\usepackage{subfig} 

\usepackage{algorithmicx}
\usepackage{algpseudocode}
\usepackage[ruled]{algorithm}
\usepackage{setspace}
\definecolor{light-gray}{gray}{0.75}
\algrenewcommand{\algorithmiccomment}[1]{\hskip3em{{\footnotesize \textcolor{light-gray}{$\blacktriangleright$}}} #1}

\usepackage{multirow} 
\usepackage{rotating} 
\usepackage{booktabs} 
\usepackage{colortbl} 
\usepackage{tablefootnote} 
\usepackage{array} 
\newcolumntype{L}[1]{>{\raggedright\let\newline\\\arraybackslash\hspace{0pt}}m{#1}}
\newcolumntype{C}[1]{>{\centering\let\newline\\\arraybackslash\hspace{0pt}}m{#1}}
\newcolumntype{R}[1]{>{\raggedleft\let\newline\\\arraybackslash\hspace{0pt}}m{#1}}

\usepackage[colorlinks=true,urlcolor=blue,citecolor=blue]{hyperref}

\usepackage{xspace}

\usepackage{enumitem}

\usepackage{bbding}
\usepackage{pifont}
\usepackage{wasysym}
\usepackage{oplotsymbl}

\usepackage{blindtext}

\newcommand{\mobilenet}{MobileNet\xspace}
\newcommand{\resnet}{ResNet\xspace}
\newcommand{\lenet}{LeNet\xspace}

\begin{document}

\title{Characterizing the Performance of Accelerated Jetson Edge Devices for Training Deep Learning Models{\thanks{~Journal article published in POMACS 2022: Prashanthi S.K, Sai Anuroop Kesanapalli, and Yogesh Simmhan, “Characterizing the Performance of Accelerated Jetson Edge Devices for Training Deep Learning Models,” in \textit{Proceedings of the ACM on Measurement and Analysis of Computing Systems, Volume 6, Issue 3, 2022 \href{https://doi.org/10.1145/3570604}{doi:10.1145/3570604}}}}} 

\author{Prashanthi S.K, Sai Anuroop Kesanapalli and Yogesh Simmhan\\
Department of Computational and Data Sciences\\
Indian Institute of Science\\
Bangalore 560012 INDIA\\
Email: \{prashanthis, simmhan\} @iisc.ac.in
}
\date{}

\maketitle

\begin{abstract}

Deep Neural Networks (DNNs) have had a significant impact on domains like autonomous vehicles and smart cities through low-latency inferencing on edge computing devices close to the data source. However, DNN training on the edge is poorly explored.
Techniques like federated learning and the growing capacity of GPU-accelerated edge devices like NVIDIA Jetson motivate the need for a holistic characterization of DNN training on the edge.
Training DNNs is resource-intensive and can stress an edge's GPU, CPU, memory and storage capacities. Edge devices also have different resources compared to workstations and servers, such as slower shared memory and diverse storage media.
Here, we perform a principled study of DNN training on individual devices of three contemporary Jetson device types: AGX Xavier, Xavier NX and Nano for three diverse DNN model--dataset combinations. We vary device and training parameters such as I/O pipelining and parallelism, storage media, mini-batch sizes and power modes, and examine their effect on CPU and GPU utilization, fetch stalls, training time, energy usage, and variability. 
Our analysis exposes several resource inter-dependencies and counter-intuitive insights, while also helping quantify known wisdom. Our rigorous study can help tune the training performance on the edge, trade-off time and energy usage on constrained devices, and even select an ideal edge hardware for a DNN workload, and, in future, extend to federated learning too. 
As an illustration, we use these results to build a simple model to predict the training time and energy per epoch for any given DNN across different power modes, with minimal additional profiling.
\end{abstract}

\section{Introduction}

\textit{Motivation.}~Deep Neural Network (DNN) models are becoming ubiquitous in a variety of 
contemporary domains 
such as Autonomous Vehicles~\cite{kuutti2020survey}, Smart cities~\cite{chen2019survey} and Healthcare~\cite{kollias2018deep}. They help drones to navigate, identify suspicious activities from safety cameras, and perform diagnostics over medical imaging. Fast DNN \textit{inferencing} close to the data source is enabled by a growing class of accelerated edge devices such as NVIDIA Jetson and Google Coral which host low-power Graphics Processing Units (GPUs) and Tensor Processing Units (TPUs) along with ARM CPUs in a compact form-factor to offer a superior performance-to-energy ratio. E.g., the NVIDIA Jetson AGX Xavier kit has a $512$-core Volta GPU, an $8$-core ARM CPU and $32GB$ of LPDDR4x memory, that operates within $65W$ of power, costs US\$999 and is smaller than a paperback novel (see Table~\ref{tbl:jetsonspecs}).

Recently, there has been a push towards \textit{training} DNN models on the edge~\cite{trainedge,edge_review,paise22}. This is driven by the massive growth in data collected from edge devices in Cyber-Physical Systems (CPS) and Internet of Things (IoT), the need to refresh the models periodically, the bandwidth constraints in moving all this data to Cloud data centers for training, and a heightened emphasis on privacy by retaining data on the edge. This has led to techniques like federated and geo-distributed learning~\cite{zhou2019edge} that 
train DNN models locally on data on an edge device and aggregate them centrally.

\textit{Gaps.}~The proliferation of DNN training has led to an increasing interest in scrutinizing the system characteristics of such workloads to help identify bottlenecks, optimize the training parameters and even to choose the machine configuration. However, this has been largely limited to evaluating their performance on GPU-accelerated Cloud VMs and servers~\cite{mohan2021analyzing, wang2019benchmarking}, with minimal investigations of edge accelerators. Profiling on the edge has been limited to inferencing workloads~\cite{baller2021deepedgebench, edge_config}. 

Accelerated edge devices like NVIDIA's Jetson series have \textit{unique characteristics} such as low-power usage, a slower RAM shared between GPU and CPU, support for diverse storage media such as SD cards, eMMC and NVME SSDs, and the ability to dynamically configure the active CPU cores and processor frequencies. So, understanding the performance characteristics and resource inter-dependencies of accelerated edge devices for DNN training is essential to design efficient DNN frameworks, optimize system resources for the constrained hardware, and schedule federated learning intelligently. This can also help in making informed design choices on configuring edge devices tailored for specific DNN training workloads, including federated learning in future.

\textit{Goals \& Outcomes.}~In this paper, we address this gap in literature by conducting a principled empirical study of three contemporary Jetson device types -- \textbf{AGX} Xavier, Xavier \textbf{NX} and \textbf{Nano} -- by training three popular DNN models using PyTorch under diverse system and training configurations. We discern the impact of hardware resources such as the number of CPU cores, CPU/GPU/memory frequency, storage media and power modes on the training time and energy usage.  We also examine how PyTorch settings such as the number of concurrent data loaders, and the DNN model and data sizes, affect the performance. These are reported as a series of ``take-aways'' for practitioners to tune their edge device and DNN framework, as well as systems researchers to help design systems software for better performing and sustainable DNN training on the edge. 

While some of our analyses confirm expected behavior with quantification, several other insights are counter-intuitive. E.g., purchasing a faster and more expensive storage may not necessarily improve the training speed if pipelining and caching are able to hide the GPU stalls; a slower and cheaper hard disk could give the same performance. Similarly, a power mode with the highest GPU frequency but a lower CPU frequency may not give benefits for smaller DNN models like \lenet which are CPU bound due to pre-processing costs. As a practical utilization of our learning, we train a simple linear-regression model to predict the expected training time and energy use for a given DNN architecture with minimal profiling information.

\emph{Non-goals.} This work focuses on training on a single Jetson edge device. We do not profile network I/O, model parallelism or model aggregation that are required for distributed training and/or federated learning. That said, characterizing the ``local model'' training on a single device is a necessary step towards analyzing distributed workloads. Our article is limited to Jetson devices as other edge accelerators like Movidius VPU and Coral TPU are too constrained for training. The profiling approach we take here can serve as a template to study future accelerated edge devices.

\textit{Contributions.}~We make the following specific contributions in this article through a methodical profiling of Jetson edge accelerators for training DNN models locally on a single device:
\begin{enumerate}
    \item We understand the effect of \textit{disk caching, pipelining and parallelizing data fetch and pre-processing} on the stall time and epoch training time, and the interplay between CPU and GPU performance (Sec.~\ref{sec:analysis:pipe}).
    \item We study the impact of \textit{storage medium and mini-batch sizes} on stalls, GPU compute time and end-to-end times (Sec.~\ref{sec:analysis:disk}, Sec.~\ref{sec:analysis:bs}), and confirm the \textit{deterministic performance} of these devices across time and instances when training (Sec.~\ref{sec:analysis:var}).
    \item We investigate the consequence of \textit{Dynamic Voltage and Frequency Scaling (DVFS) and various power modes} on the training time, energy usage and their trade-off (Sec.~\ref{sec:analysis:dvfs}, Sec.~\ref{sec:analysis:power}). 
    \item Lastly, we use these results to train simple models to predict the epoch training time and the energy usage per epoch of a given DNN for any power mode with limited profiling  (Sec.~\ref{sec:analysis:predict}).
\end{enumerate}

These are preceded by a background on edge accelerators and training (Sec.~\ref{sec:bg}), related work on characterising DNNs on various platforms (Sec.~\ref{sec:related}) and details of our experiment setup (Sec.~\ref{sec:setup}).

\section{Background and Motivation}%
\label{sec:bg}

\subsection{Edge Accelerators}
NVIDIA Jetsons have become popular as accelerated edge devices due to having similar micro-architectures as their widely-used workstation and server GPUs, albeit with fewer cores; and strong software and SDK support to build ML applications. Jetson devices are available as accelerator modules with CPU, GPU and memory for industries that build custom hardware (e.g., on self-driving cars), or as developer kits where the modules are coupled with NVIDIA's reference carrier board to form a fully working edge system for evaluation. 
These devices are becoming more powerful over time, even as they offer a low power envelope and a compact form-factor~\cite{paise22}. E.g., NVIDIA's latest edge-accelerator kit, AGX Orin, released in April 2022, delivers a theoretical $275$~TOPS of performance and features a $12$-core ARM Cortex A78AE CPU, an Ampere GPU with $2048$ CUDA cores and $64$ tensor cores, and $32GB$ of shared RAM.
These are comparable to an RTX 3080 Ti workstation GPU, but with a power consumption of $\leq 60W$ 
and no larger than a paperback novel. Therefore, these accelerators are competitive candidates for running DNN training workloads.

These edge devices have several unique features that warrant a careful study for training:
\begin{itemize}
    \item The RAM is shared between the CPU and GPU, unlike in workstation/server GPUs which have a dedicated RAM. 
    This increases the amount of memory available to the GPU, but brings in the interplay between the memory used by CPU and GPU for the model, dataset, cache, etc. 
    Also, the LPDDR RAM used in these devices is slower and low-powered, as opposed to GDDR that is used in regular GPUs.
    \item Edge devices offer several inbuilt and user-defined power modes, each with different cores, CPU frequency, GPU frequency and memory frequency. This offers a large parameter space ($>29k$ combinations for AGX) with interesting power--performance trade-offs. A close understanding of these trade-offs can help select power modes that, say, reduce over-heating of an edge by using a low-power mode while still meeting a training time budget.
    \item They support a wide variety of storage media including eMMC, Micro SD card, NVME Solid State Drive (SSD), Hard Disk Drive (HDD), which have different I/O performance and monetary costs. These can affect I/O intensive workloads like DNN training.
\end{itemize}

Two other prominent edge accelerators are \textit{Google Coral}~\cite{coral} and \textit{Intel Movidius Neural Compute Stick (NCS)}~\cite{movidius}. Coral features Google's Edge TPU accelerator for inferencing, and is available as a developer board and a USB accelerator that is connected to a host such as Raspberry Pi. Movidius uses the Intel Myriad X Vision Processing Unit (VPU) as an accelerator and is available as a USB stick. Both these devices are intended for inferencing with an extremely low power budget. They also offer limited memory. E.g., the Coral board's TPU has an on-chip memory of just $8MB$ and an off-chip memory of $1$--$4GB$, and operates within $2W$~\cite{dev_datasheet}. Additionally, Movidius's OpenVINO software development kit does not allow training. Given these constraints, it is not practical to perform DNN training on them, and we limit our study to more capable Jetson edge devices.

\subsection{DNN Training}

\begin{figure}[t]
 \vspace{-0.1in}
\centering
\includegraphics[width=0.9\textwidth]{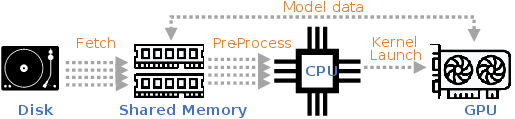}
\vspace{-0.1in}
\caption{DNN training stages using PyTorch on the Edge} \label{training-pipeline}
\vspace{-0.15in}
\end{figure}

DNN training happens iteratively (Fig.~\ref{training-pipeline}). In each iteration, we \textit{fetch} a ``mini-batch'' of samples from disk to memory, and perform \textit{pre-processing} on the mini-batch, such as deserialization, cropping, resize, flipping and normalization of the input images using the CPU. 
Then, the CPU launches \textit{kernels} on the GPU to perform the forward and backward passes of training. This repeats for the next mini-batch and so on until all the input samples are consumed. This forms one \textit{epoch} of training. Epochs are repeated using different mini-batch samplings till the model converges. 

Fetching a mini-batch from disk is I/O intensive, pre-processing is CPU intensive and the training is GPU intensive. Since the GPU is often the bottleneck during overall training, the goal is usually to maximize the GPU utilization to reduce the end-to-end training time. Performing these three stages sequentially will cause the GPU to remain idle while it waits for the disk and CPU to finish fetching and pre-processing a mini-batch. PyTorch's \texttt{DataLoader} and input pipelines constructed from TensorFlow's \texttt{tf.data} API help pipeline the fetch and pre-process stages with the compute stage. 

PyTorch has emerged as a popular DNN framework because of its ease of use as compared to TensorFlow, and more pre-trained models being available~\cite{TFvsPy}. Hence, we conduct our study using PyTorch. DNN training in PyTorch can be pipelined across the fetch and pre-processing stages, and the GPU compute stage. While the first two stages are executed sequentially by a single worker process, i.e., not pipelined, the latter can be executed in a \textit{pipelined} manner by a separate process. 
The fetch and pre-process stages can also be \textit{parallelized} to operate on multiple mini-batches so that a fast GPU does not have to wait, or ``stall'', for a mini-batch to be ready. The CPU is responsible for loading DNN kernels needed for the forward and backward pass to the GPU. If the CPU is unable to launch the kernels in time, the GPU stalls.

\section{Related Work}
\label{sec:related}

There is growing interest in training DNNs on the edge
and it offers several systems research challenges~\cite{paise22}.
However, there is a lack of rigorous empirical studies to characterize their performance and the specific challenges in effectively leveraging them. We address this gap in this paper. 

Liu et al.~\cite{tx2_training} examine DNN training workloads on the Jetson TX2 with respect to memory, CPU/GPU utilization and power consumption. They also correlate the analysis to lower level operations in DNN models. However, they do not experiment with varying power modes and framework configurations as we do, and the TX2 is an older architecture.
The Flower federated learning framework~\cite{beutel2020flower} supports heterogeneous environments including edge devices. They present results of deploying Flower on virtual Android devices and on Jetson TX2 edge accelerators. However, the edge is just a validation platform in their work and they do not offer any detailed analysis of the performance of the edge for training.

There exists literature on evaluating edge devices for model inferencing.
DeepEdgeBench~\cite{baller2021deepedgebench} compares the inference time and power consumption for edge devices such as NVIDIA Jetson Nano, Google Coral and Raspberry Pi 4 for \mobilenet v2 but training is not considered.
Others~\cite{tk1_europar} study Jetson TX1 and TK1 using roofline models for both the CPU and GPU with a matrix multiplication as the workload. While important, matrix multiplication is limited to the GPU. The training pipeline is I/O intensive and exercises disk, memory, CPU and GPU. We study this holistically.

MLPerf~\cite{MLSYS2020_02522a2b} is a community effort to provide a uniform framework for quantifying the performance of ML Hardware and Systems. The benchmark suite spans a number of application domains and datasets, and prescribes a quality threshold that must be met by any implementation. While such a suite is essential for measuring the overall impact of systems or optimizations on training, it does not measure low-level system metrics like the IO reads. MLPerf also lacks a training suite for edge devices. We adopt a similar training benchmark in our study, which is viable on edge devices.

There has also been specific attention on the energy usage and variability of edge devices. Holly, et al.~\cite{edge_config} correlate CPU and GPU frequencies and the number of CPU cores with the latency, power and energy for inferening on the Jetson Nano. Some~\cite{frqswitching} examine the effect of these on power consumption for stream processing workloads. We focus on the impact of such configurations, including storage media and DNN framework settings, on the end-to-end time and energy consumed for DNN training workloads on the Jetson AGX Xavier. Snowflakes~\cite{snowflakes} reports a detailed study on the latency and power variability observed across Jetson AGX Xaviers for inferencing. We too evaluate the variability, but for a training workload and we do not observe any variability.

There is a larger body of work on training on GPU workstations and servers.
Mohan, et al.~\cite{mohan2021analyzing} characterize the training data pipeline and how it affects training time on desktop GPUs. They also analyze the effect of the OS page cache on data access. However, their study only considers server-grade GPUs which are much more powerful and have exclusive and faster GPU RAM when compared to edge devices.
They also propose a modified caching mechanism that minimises I/O caused by thrashing. Once the page cache is full, all further accesses are sent to disk without evicting existing data in the cache. Quiver~\cite{kumar2020quiver} proposes a caching strategy based on substitutability. Accesses that cause a miss in the cache are substituted with other data that are present in the cache without interfering with training requirements and single access per epoch.
Our detailed study and analysis can help design such optimization strategies for training on edge accelerators.

Lastly, our paper enables accurate modeling of DNN training time and energy usage for the diverse power modes of these devices. This is key for federated learning when devices in a training round need to complete at about the same time~\cite{google-sysml}. Current techniques use simple approximations like over-sampling of devices~\cite{google-sysml}. We provide initial promising results in this direction.

\begin{table}[t]
\centering
\footnotesize
\vspace{-0.1in}
\caption{Specifications of NVIDIA Jetson Devices Evaluated}
\label{tbl:jetsonspecs}
\vspace{-0.1in}
\begin{tabular}{L{3.5cm}|R{1.5cm}|R{2.3cm}|R{2.3cm}|R{2.3cm}}
\toprule
\textbf{Feature} & \textbf{Nano}~\cite{powermodes_nano}  &  Xavier \textbf{NX}~\cite{powermodes_nxagx} & \textbf{AGX} Xavier~\cite{powermodes_nxagx} & AGX \textbf{Orin}~\cite{powermodes_orin}\\
\midrule
ARM CPU Architecture & A57 & Carmel & Carmel &  A78AE\\ 
\noalign{\global\arrayrulewidth=0.1pt}\arrayrulecolor{lightgray}\hline
\noalign{\global\arrayrulewidth=0.4pt}\arrayrulecolor{black}
CPU Cores$^\dagger$ & 4 & 6 & 8 & 12\\
\noalign{\global\arrayrulewidth=0.1pt}\arrayrulecolor{lightgray}\hline
\noalign{\global\arrayrulewidth=0.4pt}\arrayrulecolor{black}
CPU Frequency (MHz)$^\dagger$ & 1479 & 1900 & 2265 & 2200\\ \hline
GPU Architecture & Maxwell & Volta & Volta & Ampere\\
\noalign{\global\arrayrulewidth=0.1pt}\arrayrulecolor{lightgray}\hline
\noalign{\global\arrayrulewidth=0.4pt}\arrayrulecolor{black}
CUDA/Tensor Cores & 128/--& 384/48 & 512/64 & 2048/64\\
\noalign{\global\arrayrulewidth=0.1pt}\arrayrulecolor{lightgray}\hline
\noalign{\global\arrayrulewidth=0.4pt}\arrayrulecolor{black}
GPU Frequency (MHz)$^\dagger$ & 921 & 1100 & 1377 & 1300\\ \hline
RAM (GB) & 4 & 8 & 32 & 32\\ \hline
Storage Interfaces & $\mu$SD, USB & $\mu$SD, NVMe, USB & $\mu$SD, eMMC, eSATA, NVMe, USB & $\mu$SD, eMMC, NVMe, USB \\
\noalign{\global\arrayrulewidth=0.1pt}\arrayrulecolor{lightgray}\hline
\noalign{\global\arrayrulewidth=0.4pt}\arrayrulecolor{black}
Memory Frequency (MHz)$^\dagger$ & 1600 & 1600 &  2133 & 3200\\ \hline
Power (W)$^\dagger$ & 10 & 15$^*$ & 65$^\#$ & 60\\ \hline
Price (USD) & $\$129$ & $\$399$ & $\$999$ & $\$1999$  \\
\bottomrule
\multicolumn{5}{L{13cm}}{$^\dagger$~This is the maximum possible value across all power modes.
Actual value depends on the power mode used (Table~\ref{tbl:power}).
\quad 
$^*$ This peak power is for Jetpack release $v4.5.1$ and earlier. 
\quad 
$^\#$ The data sheet does not list the power for the MAXN peak power mode. We report the power adapter rating of $65W$.
}
\end{tabular}
\vspace{-0.15in}
\end{table}

\section{Experiment Setup}
\label{sec:setup}

\subsection{Hardware Platform}

We perform our experiments on three contemporary classes of NVIDIA Jetson developer kits:  \textbf{AGX} Xavier~\cite{AGX}, Xavier \textbf{NX}~\cite{NX} and \textbf{Nano}~\cite{Nano} (for convenience, we refer to the devices by the names highlighted in bold). We use five devices of each type in our experiments. \textbf{Orin} AGX~\cite{Orin}, released in April 2022, was available in the market just recently and we offer some early results on it. The specifications of the devices are given in Table~\ref{tbl:jetsonspecs}. 

Briefly, the \textit{Nano} has $4$ ARM A57 CPU cores at a peak frequency of $1.479GHz$ 
and a Maxwell GPU with $128$ CUDA cores at a peak frequency of $921MHz$. It has $4GB$ of shared LPDDR RAM -- a low-power but slower variant of GDDR -- a peak power of $10W$ and costs US\$~$129$. 
The \textit{NX}, a more powerful variant, comes with 6 Carmel cores at a peak frequency of $1.9GHz$ in dual-core mode and a Volta GPU with $384$ CUDA cores and $48$ tensor cores. It has $8GB$ of shared LPDDR RAM, a peak power of $15W$ and costs US\$~$399$. 
The \textit{AGX} uses NVIDIA's custom Carmel ARM CPU with $8$ cores, along with a Volta GPU with $512$ CUDA cores. It has $32GB$ of shared LPDDR RAM and a list price of US\$~$999$.
\textit{Orin} features a $12$-core ARM Cortex A78AE, an Ampere GPU with $2048$ CUDA cores and $64$ tensor cores and $32GB$ 
of shared  LPDDR RAM, with a peak power of $60W$. It sells for US\$~$1999$.
The RAM is shared between CPU and GPU on all these classes of devices.

These edge devices offer interfaces to different storage media. 
The Nano supports a Micro SD card and a USB HDD. The NX supports a Micro SD card, USB HDD and M.2 NVME SSD. The AGX and Orin come with an eMMC (flash based) storage and support a Micro SD card, HDD over USB, and M.2 NVME SSD. AGX also supports HDD over eSATA.

In our experiments, the OS and platform binaries are installed on the eMMC for the AGX and Orin, and on the Micro SD card for the NX and Nano.
Since training can be I/O intensive, we perform experiments by hosting the training data on three different storage media -- \textit{SSD over an NVMe/PCIe interface}, \textit{HDD over a USB 3.0 interface}, and \textit{SD card}.
We use a $64GB$ Samsung EVO Plus Micro SD card, a $250GB$ M.2 NVME Samsung SSD 980, and a $1TB$ Western Digital My Passport HDD over USB. The peak sequential read speeds from their datasheets are $3.5GBps$ for the SSD, $1.05GBps$ for the HDD, and $100MBps$ for the SD card. 

\vspace{-1ex}
\subsection{Software Platform}
All $15$ devices of the AGX, NX and Nano run Linux for Tegra (L4T) $v32.5.1$ with $v4.9.201$-tegra kernel. They have CUDA $v10.2$ with Jetpack $v4.5.1$. We use PyTorch $v1.8$ and Torchvision $v0.9$ as the DNN training framework. However, Orin requires a more recent OS and library version: CUDA $v11.2$, Jetpack $v5.0.1$ running on L4T $v34.1.1$, Pytorch $v1.12$ and Torchvision $v0.13$.

We use the PyTorch framework for training~\cite{pytorch} with the \texttt{Data\-loader}~\cite{dataloader} to fetch and pre-process data. 
We use the \texttt{num\_workers} flag to vary the number of fetch and pre-process workers. 
When \texttt{num\_workers=0}, a single process performs fetch, pre-process \textit{and} GPU compute sequentially, without pipelining. When \texttt{num\_workers $\geq$ 1}, PyTorch spins up that many processes for fetch/pre-process, each operating on a different batch of data in parallel, and a separate process invokes the GPU compute on each pre-processed batch sequentially. This forms a two-stage pipeline of fetch/pre-process followed by compute. 

\vspace{-1ex}
\subsection{DNN Models and Datasets}

\begin{table*}[t]
\centering
\setlength{\tabcolsep}{1pt}
\def\arraystretch{0.95}
\footnotesize
\vspace{-0.1in}
\caption{DNN Models and Training Datasets}
\label{tbl:modeldataset}
\vspace{-0.1in}
\begin{tabular}
{lrrR{1.4cm}r|lR{1.5cm}R{1.2cm}R{0.9cm}}
\toprule
 \bf{Model} & \bf{\# Layers} & \bf{\# Params} & \bf{Mem. Used}~\cite{summary} & \bf{FLOPs} & \bf{Dataset} & \bf{\# Train. Samples} & \bf{Size on Disk} & \bf{Batch Size}\\
  \midrule
 \textbf{\lenet -5} & 7 \cite{LeNet} &  $60k$ \cite{LeNet} & $0.35MB$ &  $4.4M$ \cite{ding2018auto}  &  \textbf{MNIST} &  $60,000$ \cite{LeNet}  & $46MB$ &  $16$\\
  \hline
 \textbf{{\mobilenet}\,v3}& 20 \cite{howard2019searching} & $5.48M$ \cite{mobnet_flops} & $124.73MB$& 
$225.4M \cite{mobnet_flops}$
 & \textbf{GLD23k} & $23,080$ \cite{tensorflow_gld23k} & $2827MB$ 
 & $16$\\
  \hline
 \textbf{\resnet -18}& 18 \cite{he2016deep} & $11.68M$ \cite{flops} & $53.89MB$ & 
$1.82G$   \cite{flops}
 & \textbf{CIFAR-10} & $50,000$ \cite{krizhevsky2009learning}& $150MB$ & $16$\\
 \midrule
 \textbf{VGG-11}& 11 \cite{vgg} & $132.86M$ \cite{flops} & $509.78MB$ & 
$7.63G$   \cite{flops}
 & \textbf{CIFAR-10} & {---''---}& {---''---} & {---''---}\\
\bottomrule
\end{tabular}
\vspace{-0.15in}
\end{table*}

\begin{table*}[t]
\centering
\setlength{\tabcolsep}{1pt}
\caption{DNN Models and Device Trained On}
\label{tbl:sigmetricsmodeldataset}
\begin{tabular}
{L{3cm}|ccc|c}
\toprule
 \bf{Model} & \bf{AGX} & \bf{NX} & \bf{Nano} & \bf{Orin}\\
  \midrule
 \lenet -5 &\checkmark & \checkmark & \checkmark & \checkmark\\
  \hline
 \mobilenet -v3 &\checkmark & \checkmark & & \checkmark\\
  \hline
 \resnet -18& \checkmark & &  & \checkmark\\
 \hline
 \resnet -50&  & & & \checkmark\\
 \hline
 VGG-11 & \checkmark & &  \\
 \hline
\end{tabular}
\end{table*}

We chose three DNN models for computer vision for our training experiments -- \lenet-5, \mobilenet v3 and \resnet-18. This was based on their popularity observed from our survey of around $60$ edge and federated learning research papers. These provide a variety of DNN architectures and computational footprints, as shown in Table~\ref{tbl:modeldataset}. 

We investigated the MLPerf community benchmark as an evaluation suite~\cite{MLSYS2020_02522a2b}. However, they do not have workloads for edge training but only for edge inferencing and for workstation/server training. Additionally, the benchmark only reports coarse-grained metrics such as inference latency and time-to-train, but not other metrics such as stall time, compute time, IOPS, etc. We pick similar but smaller DNN models as the vision area of MLPerf, e.g., ResNet-18 instead of ResNet-50.

\textit{\lenet} is one of the earliest and simplest Convolutional Neural Network (CNN) models designed to recognize handwritten digits, $0$--$9$, for Optical Character Recognition (OCR). 
We train the \lenet -5 DNN~\cite{LeNet} on the \textit{MNIST}~\cite{LeNet} dataset. It consists of $60,000$ training and $10,000$ test images, each of which is a $28\times28$ grayscale image of a handwritten digit in the class $0$--$9$. Google's \textit{\mobilenet}~\cite{howard2019searching} is a lightweight model intended 
for vision-based applications on mobile devices.
We use images from the \textit{Google Landmarks Dataset v2 (GLD-23k)}~\cite{weyand2020google} to train \mobilenet-v3 over $23,080$ images of human-made and natural landmarks, divided into $203$ classes, with a total size on disk of $2.8~GB$. \textit{Residual Neural Network (\resnet)} is a class of CNNs that are used for vision-based applications. \textit{CIFAR-10}~\cite{krizhevsky2009learning} is used to train the \resnet-18 DNN. It has $50,000$ training and $10,000$ test images. The size of the training files with $50k$ images is $150MB$. Additionally, we use \textit{VGG11}~\cite{vgg}, a popular CNN, with the \textit{CIFAR-10} dataset to validate the epoch-time prediction model we train in Sec.~\ref{sec:analysis:predict}.
Not all models fit within the available memory of all edge devices.  Each model is trained only on the devices that have sufficient memory for training, as indicated in Table~\ref{tbl:modeldataset}. For instance, \resnet-18 is trained only on the AGX because both the NX and the Nano run out of memory. Since AGX supports all models and datasets evaluated, and it is a newer hardware platform, some of our experiments drill-down into the AGX as a canonical edge accelerator.

\vspace{-1ex}
\subsection{Default Configuration}
\label{sec:setup:default}
We use the following default configurations in our experiments based on best practices from literature~\cite{snowflakes,frqswitching}, unless stated otherwise. 
The default power mode is the highest rated for all devices: MAXN for the AGX and Nano, and $15W$ for NX 
(modes $g$, $MAXN$ and $15W$ in Table~\ref{tbl:power}). DVFS is turned off. The fan speed is set to maximum to avoid resource throttling due to overheating. By default, we store the training data on SSD for the AGX and NX, and on SD card for the Nano. In experiments where we need the same storage media type across all three device classes, we use HDD over USB for the training data as it is present on all.

The prefetch factor in PyTorch \texttt{DataLoader} is set to its default value of $2$. The number of worker processes in the \texttt{DataLoader} is set to $w=4$, for reasons discussed in Sec.~\ref{sec:analysis:pipe}. Previous works ~\cite{golmant2018computational,charles2021large,shallue2018measuring} have shown that large mini-batch sizes adversely affect convergence and therefore we use a mini-batch size of $bs=16$ images when training, as this is a small mini-batch size commonly used across models. The learning rate and momentum are set to $0.01$ and $0.9$ respectively~\cite{chen2022demon}.
We use Stochastic Gradient Descent (SGD) as the optimizer, and cross-entropy as the loss function.
We clear the page cache at the start of every experiment run to avoid any cross-experiment effects, but it is retained across epochs within a single training run.

In each experiment, we train the DNN models for $6$ epochs. As we show in Section~\ref{sec:analysis:var} for a $15h$ run, this is adequate to understand and generalize the performance behavior when training till convergence. 
Also, for some configurations, each epoch takes $90~mins$. We do not include a testing phase in our experiments as we are not training till convergence. 
By default, we report the results averaged over epochs $1$--$5$ since epoch $0$ has bootstrapping overheads, as discussed in Sec.~\ref{sec:analysis:pipe}.

\vspace{-1ex}
\subsection{Performance Metrics}
We use a variety of Linux system utilities to monitor and report system resource usage. \textit{CPU, GPU and RAM utilization}, and \textit{average and instantaneous power} are measured using the \texttt{jtop} Python module, which internally uses the \texttt{tegrastats}~\cite{tegrastats} utility from NVIDIA, at $\approx 1~s$ sampling. The power measurements are from on-board sensors in the Jetsons, which capture the power load from the module but not the carrier board and peripherals. The socket load can be captured by using an external power monitor, which we use for baseload studies. However, the bulk of the variation in the energy usage during training is from the module load. So the module load reported by the on-board sensors are used in our analysis, unless noted otherwise.

The sampling interval deviates by up to $200~ms$ due to delays introduced by the rest of the monitoring harness, e.g., \texttt{iostat} takes $1s$ when run periodically.
So the \textit{total energy} for training in a duration $T$ is calculated as a sum of the instantaneous power ($p_{t_i}$ in watts) measured at time $t_i$, weighted by the duration between successive samples ($t_i - t_{i-1}$), given as $\sum_{t_i \in T} \big( p_{t_i} \cdot (t_i - t_{i-1}) \big)$. The \textit{read IOPS} and \textit{bytes read per second (throughput)} are measured using \texttt{iostat}~\cite{iostat}. The fraction of the dataset that is present in the Linux (in-memory) disk cache is measured using \texttt{vmtouch}~\cite{vmtouch}.

Additionally, we measure the fetch stall time and the GPU compute time for every mini-batch. \textit{Fetch stall time} is the \emph{visible} time taken to fetch and pre-process data, and does not overlap with the GPU compute time, i.e., $\max((\text{\textit{fetch time}} + \text{\textit{pre-process time}} - \text{\textit{GPU compute time}}), 0)$. \textit{GPU compute time} is the time taken by the mini-batch to execute the training on the GPU. It includes the kernel launch time, and the forward and backward passes of training.
We measure these times using the \texttt{torch.cuda.event} with the \texttt{synchronize} option so that time captured is accurate~\cite{Cuda_event}.

We sum the fetch stall and GPU compute times over all mini-batches in an epoch to obtain their \textit{average time per epoch}.
We also measure and report the \textit{End-to-End (E2E) time} to process all mini-batches of each epoch, including the fetch stall time, GPU compute time and any framework overheads.
We have performed multiple runs for the different experiments and they are reproducible. We report results from a representative run.

\section{Results and Analysis}
\label{sec:analysis}

We attempt to understand the impact of various hardware resource choices, hardware and OS configurations, and training platform configurations on the time taken and energy consumed for training the candidate DNN models on the edge devices. 
Specifically, we examine the impact of worker parallelism and disk caching on the I/O, pre-process and compute pipeline (Section~\ref{sec:analysis:pipe}); 
the effect of storage media on the training time (Section~\ref{sec:analysis:disk}); 
the effect of mini-batch sizes (Section~\ref{sec:analysis:bs});
the variability in training time across epochs and devices (Section~\ref{sec:analysis:var}); 
and the impact of power modes on the training time and energy usage (Section~\ref{sec:analysis:power}). Besides offering a holistic characterization of DNN training on edge accelerators, it also assists ML developers to improve the training performance by choosing the right hardware and platform setup.
Overall, we perform $\approx{5170}$ training epochs using different configurations to report our results. The scripts and logs for these are available at  
{\Note{\url{https://github.com/dream-lab/edge-train-bench/tree/sigmetrics-2023}}}. Further, we use these experimental results to develop a prediction model for the expected DNN training time per epoch and the energy per epoch for any given power mode (Section~\ref{sec:analysis:predict}). This can be used by developers of new DNNs to define a custom power mode from among, e.g., $\big($ CPU core counts (8) $\times$ CPU frequencies (29) $\times$ GPU frequencies (14) $\times$ EMC frequencies (9) $\big)$ = $29,232$ possible combinations for AGX, with a suitable time--energy trade-off with minimal prior benchmarking. 

\vspace{-1ex}
\subsection{Pipelined Training and Disk Caching}
\label{sec:analysis:pipe}

\textbf{Number of Fetch/Pre-process Workers.} 
The PyTorch dataloader lets users specify zero or more \textit{workers} ($w$) that are each responsible for fetching and pre-processing one mini-batch, and these processes \textit{pipeline} into a single GPU compute process that executes the training kernel on the GPU for this pre-processed mini-batch. Setting $w=0$ (default) causes all three stages to be executed sequentially by a single process, without any pipelining. With $w>1$, multiple workers perform the fetch and pre-process \textit{in parallel} to get batches ready for a separate GPU compute process.

Intuitively, pipelining and parallelism should reduce the training time. But the benefit varies a lot across the devices and models.
We evaluate the effect of: (1) disabling ($w=0$) and enabling ($w>0$) pipelining, and (2) the degree of parallelism of the fetch and pre-process workers ($w=\{1, 2, 4, 8\}$) 
on the training time and on the stall time for the three DNN models when running on the AGX, NX and Nano. The training data is on the HDD connected over USB for uniformity. 

\begin{figure*}[t]
\vspace{-0.15in}
\centering

  \subfloat[E2E time for Epoch $1+$, with caching and parallel workers $w=0,1,2,4$.
  ]{
    \includegraphics[width=0.85\textwidth,valign=b]{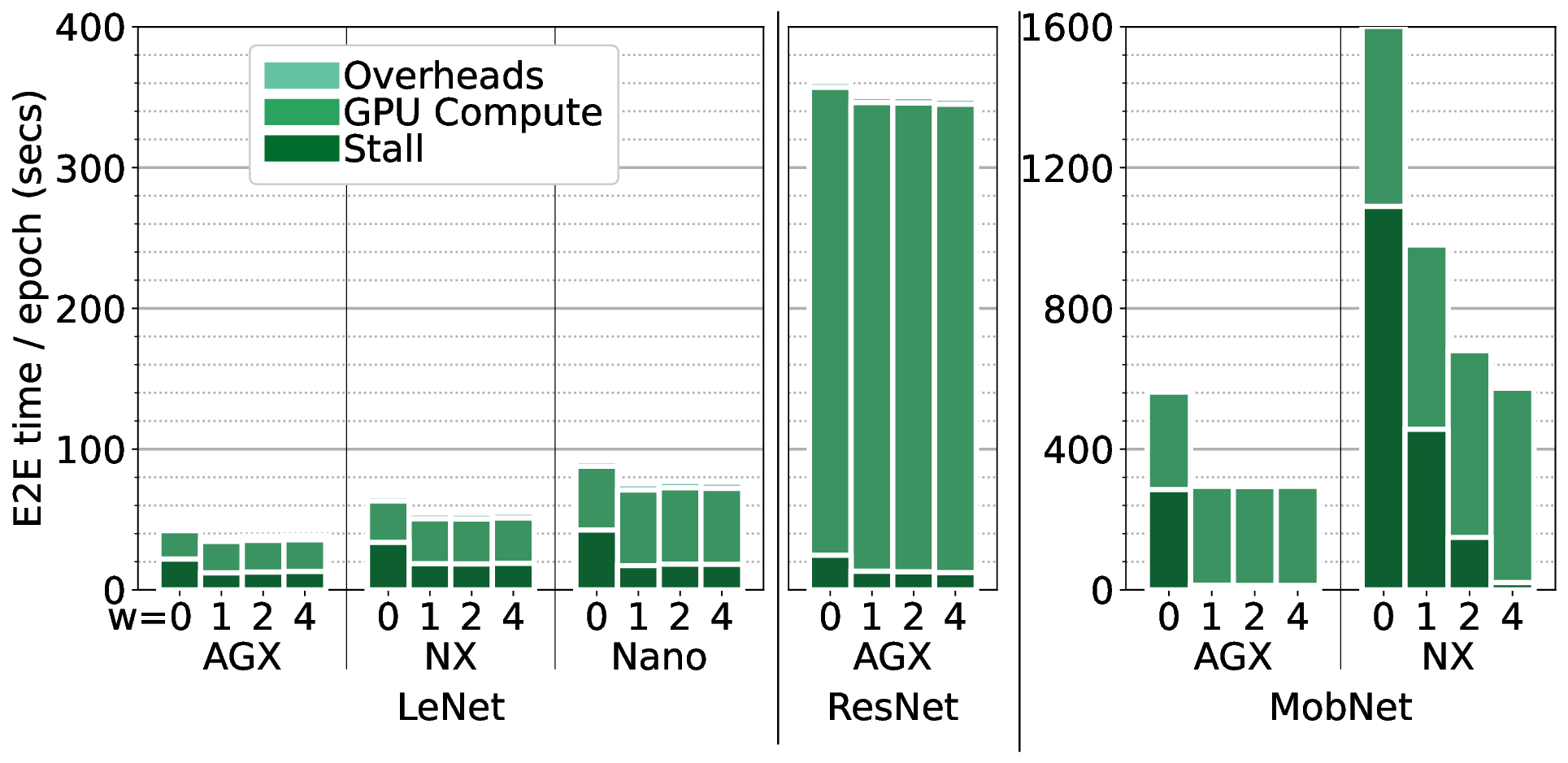}
    \label{fig:pipe:stackede2e:e1}
  }\\
  \subfloat[Stall times for Epoch $0$ (no pipelining) and Epoch $1+$, with $w=4$. ]{
    \includegraphics[width=0.45\textwidth,valign=b]{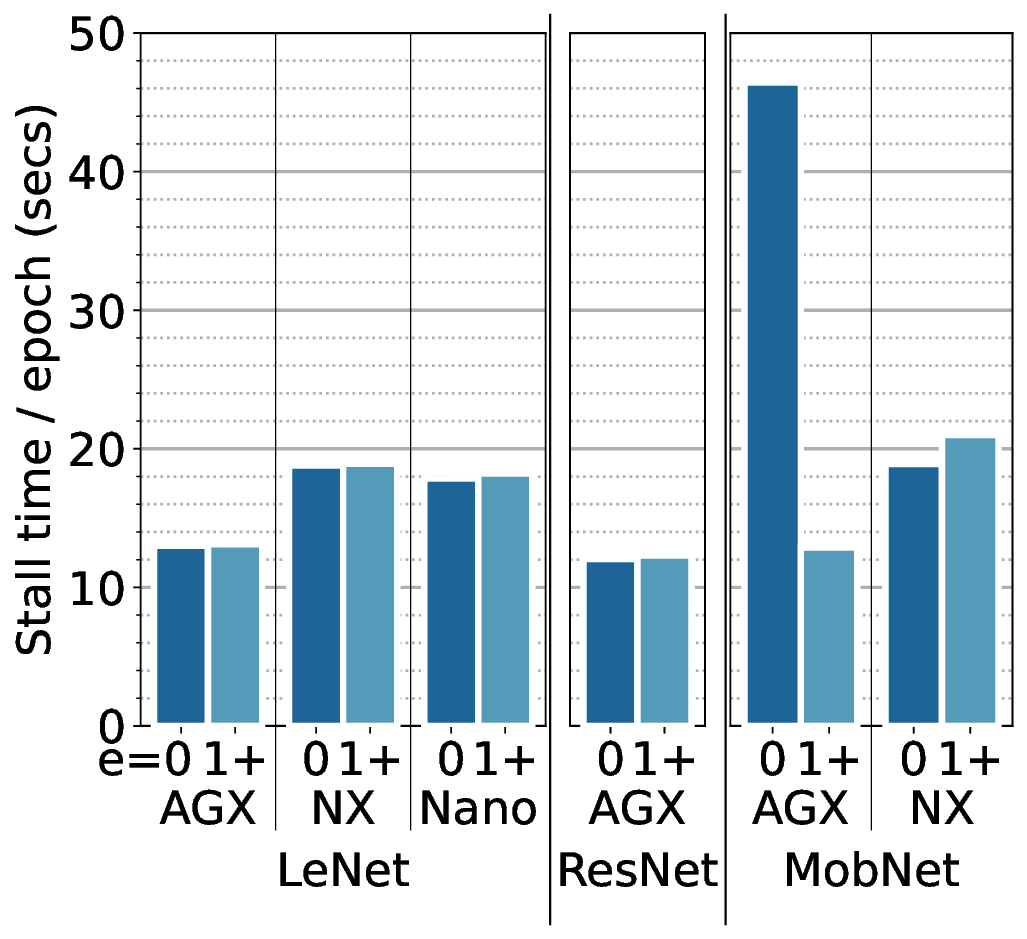}
    \label{fig:pipe:stall:e0_1+}
     }~~
  \subfloat[Energy values for Epoch $0$ with $w=4$ and Epoch $1$ with $w=0,4$.
]{
    \includegraphics[width=0.45\textwidth,valign=b]{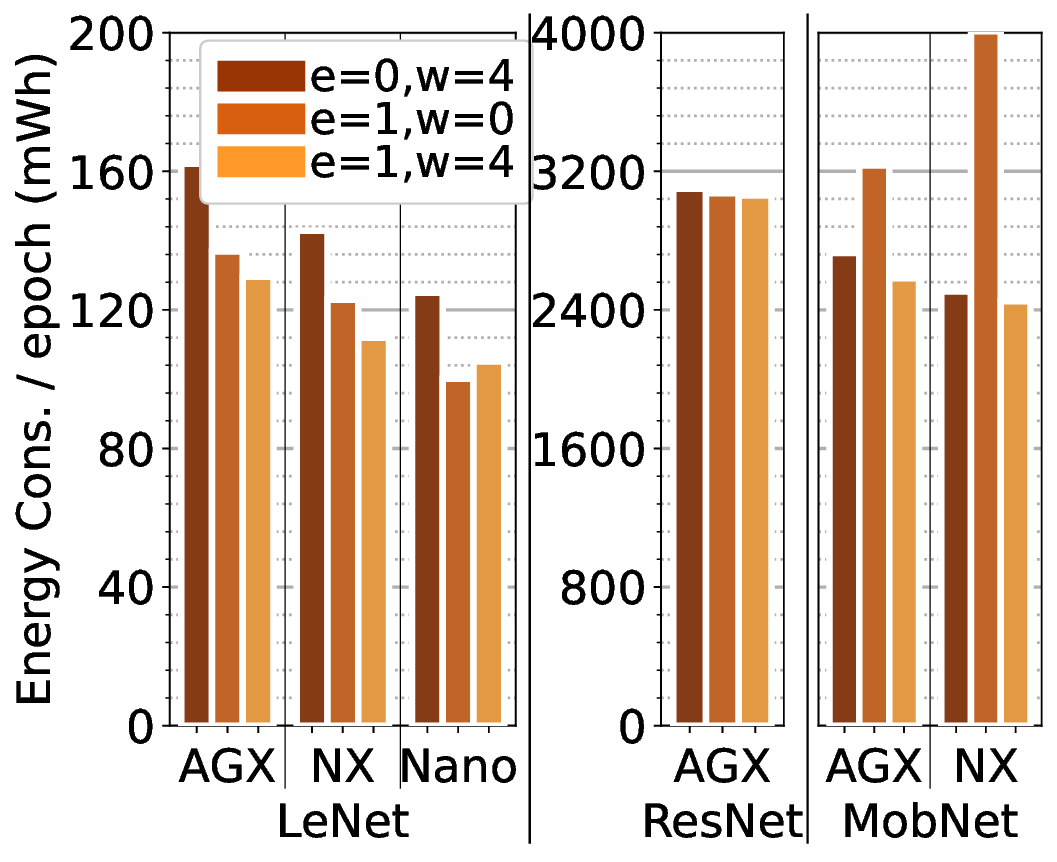}
    \label{fig:pipe:energy}
     }

  \caption{Effect of \textit{pipelining, parallel workers (w)} and \textit{disk caching} on \textit{stall time}, \textit{E2E time} and \emph{energy per epoch}.}
\label{fig:pipe:stackede2e}
\end{figure*}

Fig.~\ref{fig:pipe:stackede2e:e1} shows the total \textit{end-to-end (E2E) training time per epoch} for these configurations, and its component times -- the total fetch \textit{stall time} when the GPU was idle, waiting for a mini-batch to be ready after fetch and pre-process; the \textit{GPU compute time} where the training was happening, potentially overlapping with the fetch and pre-process stages; and the remaining \textit{overhead time} for the epoch. As discussed next, epoch $0$ has boot-strap overheads; so we exclude epoch $0$ and report an average over only epochs $1$ to $5$ ($1+$) in Fig.~\ref{fig:pipe:stackede2e:e1}. 

\textbf{Disk Caching.} The Linux \emph{page cache} uses available free memory to retain recently fetched file pages in memory. So some of the training data used in previous epoch(s) may be available in the cache for future epochs, reducing disk access. We study the effect of caching on the stall time.

At the start of every training run, we always \textit{drop the page cache} to avoid inter-experiment cache effects. This is also mimics an end-user training scenario. So, for epoch $0$, all training data will be accessed from the disk, whereas for epochs $1+$, a subset of the data may be present in and fetched from RAM, depending on the memory pressure from applications and the Least Recently Used (LRU) cache eviction policy~\cite{linux-page-lru}. To measure the impact of such caching, we report the stall times for epoch $0$ and averaged over epochs $1+$ separately in Fig.~\ref{fig:pipe:stall:e0_1+}. A typical DNN training will run for $100s$ of epochs. So epoch $1+$ is representative and epoch $0$ runs just once. Our analysis follows next.

\claim{A large page cache (RAM) can reduce the stall time.}

If \textit{all the training data} can fit within the Linux page cache, then the disk I/O is seen only for epoch $0$ where the cache is initially populated and the I/O time is eliminated for epochs $1+$. Since training typically runs over $10$--$100s$ of epochs, this can give a tangible benefit. 

In Fig~\ref{fig:pipe:stall:e0_1+}, the stall time for \mobilenet on AGX using four workers drops from $46.4s$ for epoch $0$ and to $12.9s$ for epoch $1+$. AGX's $32~GB$ of RAM is able to fit the \mobilenet model and its entire GLD dataset, which is $2.8~GB$ on disk, at the end of epoch $0$. \texttt{vmtouch} reports that $100\%$ of the training file is cached. Hence, IOPS for epochs $1+$ drops to zero and
future accesses to the training data is only from the cache.

In contrast, the stall times for \mobilenet on NX are $18.9s$ for epoch $0$ and a comparable $20.9s$ for epoch $1+$ (Fig~\ref{fig:pipe:stall:e0_1+}). NX only has $8~GB$ of RAM which is shared between GPU and CPU. The memory taken by the larger DNN model in GPU makes the available cache inadequate to retain the full training data and less than $6\%$ of data is cached as per \texttt{vmtouch}.
Since the samples in a mini-batch are randomized in each epoch, there is no data locality and no reuse of partially cached data. Linux cache's LRU policy is also not well-suited for this access pattern~\cite{mohan2021analyzing}. 
In this scenario, all accesses to training data in epochs $1+$ hit the disk, causing stalls similar to epoch $0$.

We confirm this benefit for training \mobilenet on AGX by explicitly dropping the cache after each epoch and notice that the stall times increases back to to $49.91s$ for epochs $1+$. On the other hand, there is no difference in the stall times for \mobilenet on NX even with an explicit cache-drop since the partially cached data is not reused due to lack of data locality across epochs.

\claim{A slower CPU or a smaller training data can mitigate the benefits of caching on stall time.}

Caching has limited impact when the stall time is dominated by the pre-processing stage (due to a relatively slower CPU) rather than the fetch time. Fetches can be faster due to a smaller training data or, as we will see in Sec~\ref{sec:analysis:disk}, a faster disk.

We see this happen for \lenet and \resnet on all devices. There is no visible effect of caching and the stalls are similar across all epochs. Both MNIST and CIFAR10 are small datasets and the I/O time even on a slower HDD is negligible. The IOPS drop to near-zero for epoch $1+$.
However, the time to pre-process them still takes, say $12s$ for \resnet, and this contributes to the stall time.

So, \textit{we get caching benefits only in a sweet spot}, when: (1) the training data is small enough to fully fit in the cache, i.e., within the available RAM after loading the DNN model, and yet (2) the training data also is large enough that the fetch I/O time dominates over the pre-processing time.

\claim{Pipelining reduces the stall time.}

When pipelining is enabled by increasing the workers from $w=0$ to $w=1$ for epochs $1+$, the stall time per epoch sharply reduces.
In Fig.~\ref{fig:pipe:stackede2e:e1}, the stall time (dark green stack at the bottom) for training \lenet reduces on AGX from $22s$ for $w=0$ to $12.1s$ for $w=1$, and on NX from $33.8s$ to $18.7s$.
On the Nano, the drop is even more prominent at $2.5\times$. 
Similarly, pipelining of \resnet on AGX reduces the stall time by $1.9\times$, 
while for \mobilenet on NX it drops by $2.4\times$. 
These drops are expected since pipelining hides the GPU stalls due to disk I/O and CPU pre-processing. 
AGX has a steep reduction for \mobilenet, but this is due to caching -- as seen above, caching does not benefit the other models and devices and their gains are due to pipelining.

\claim{The relative drop in E2E time due to pipelining depends on the model and the device.}
\resnet is trained on CIFAR, which uses small-sized images and has lesser I/O. However, training the \resnet model is GPU intensive. So the stall time is a small fraction of the E2E time ($6.7\%$ for epoch $0$), and pipelining reduces the overall time by only $2.9\%$. For \lenet, the model itself is light-weight and despite MNIST having a small image size, the stall time is a larger fraction of the E2E time. So the benefits of pipelining in reducing the overall epoch time is higher, giving an average of $17.7\%$ benefit across devices. \mobilenet requires modest GPU computation but the GLD images are relatively larger, resulting in significant I/O and CPU compute. Hence, pipelining halves the E2E time on AGX  
and reduces it by $38.7\%$ for NX. 

\claim{The stall time and its reduction due to pipelining are decided by the relative speeds of CPU, GPU and disk}

With pipelining enabled ($w=1$), a stall is avoided when the sum of the fetch time from disk (or cache) and pre-processing time on CPU is smaller than or comparable to the GPU compute time for a mini-batch. 
So the relative speeds of the disk, CPU and GPU, the size of the input data (which affects fetch and pre-process times), and the complexity of the DNN (which affects the GPU compute time), together determine the stall time. 

We discuss disk speed effects later in Sec.~\ref{sec:analysis:disk}.
Here, all devices use the same HDD type but have different CPU frequencies. So, as the CPU speed of a device decreases, the stall time per epoch for \lenet without pipelining ($w=0$) increases from AGX to NX to Nano, from $22s$ to $33.8s$ to $42.6s$. 
With pipelining ($w=1$), the stall times for all three devices are similar at $12.1s$, $18.7s$ and $17.2s$, respectively. 
This is because the GPU compute times are the highest for Nano, followed by the NX and the AGX, due to the increasing GPU speeds. As a result, a longer fetch and pre-process time is hidden by a similarly longer GPU compute time. So, when configuring a device, \textit{the relative speeds of the resources are more important than the absolute speed of any one resource}.

When training \mobilenet on AGX, caching eliminates the fetch time and the CPU is fast enough to pre-process the mini-batch before the GPU finishes training a prior mini-batch. So, pipelining largely hides the stall time. 

\claim{Parallelizing the fetch and pre-process may give benefits beyond pipelining.}
When we increase the number of workers to $w>1$, we can fetch and pre-process multiple mini-batches in parallel. This may further reduce the stall time, but is not guaranteed.
In Fig.~\ref{fig:pipe:stackede2e:e1}, the stall times per epoch for both \lenet and \resnet are very low ($<20s$) with $w=1$, which corresponds to $<6ms$ per mini-batch. This cannot be further reduced. For \mobilenet on AGX, the stalls are completely hidden by pipelining using $w=1$. So, an increase in $w$ does not give additional benefits in these cases.

However, for \mobilenet on the NX, we see stall times of the order of $456s$ even with pipelining with $w=1$. This is caused by the higher fetch and pre-process costs for the larger sized data, relative to the GPU compute for training the model. So, increasing $w$ from $1$ to $2$ reduces the stall time by $67.3\%$ and the E2E time by $30.5\%$ due to better disk and CPU utilization that we observe with the parallel workers.
This improvement continues as we increase $w$ to $4$ and saturates beyond $w=8$, both of which have a small stall time of $\approx 21s$ per epoch relative to an E2E time of $\approx 576s$.
\textit{Hence, we use $w=4$ workers as the default in our experiments.} 
\claim{Pipelining can reduce the energy consumption for training}

Pipelining increases the instantaneous power load across all models and devices, but the total energy consumed for the epoch is the same or lower (Fig.~\ref{fig:pipe:energy}). As fetch stalls reduce, the GPU and the CPU are utilized better and this increases the power load. However, this is offset by a drop in the training time for the epoch due to pipelining. In Fig.~\ref{fig:pipe:energy}, in going from $w=0$ to $w=4$ for \lenet epoch $1+$, the energy per epoch drops by $5.4\%$ for AGX and $8.9\%$ for NX, and increases by $4.7\%$ for Nano.
\mobilenet has a higher energy reduction of $20.1\%$ and $38.9\%$ for AGX and NX due to a larger drop in E2E time due to pipelining.

\vspace{1ex}

As a separate note, preliminary experiments on Orin show a reduction in epoch $1+$ E2E training time of $1.8$--$3.9\times$ compared to AGX, for the three DNN models. This loosely matches its $4\times$ increase in CUDA cores.
But these require further detailed investigations as future work.

\begin{figure*}[t]
\vspace{-0.15in}
 \centering
  \subfloat[Epoch $0$ (no caching), $w=0$ (no pipelining)]{
    \includegraphics[width=0.49\textwidth]{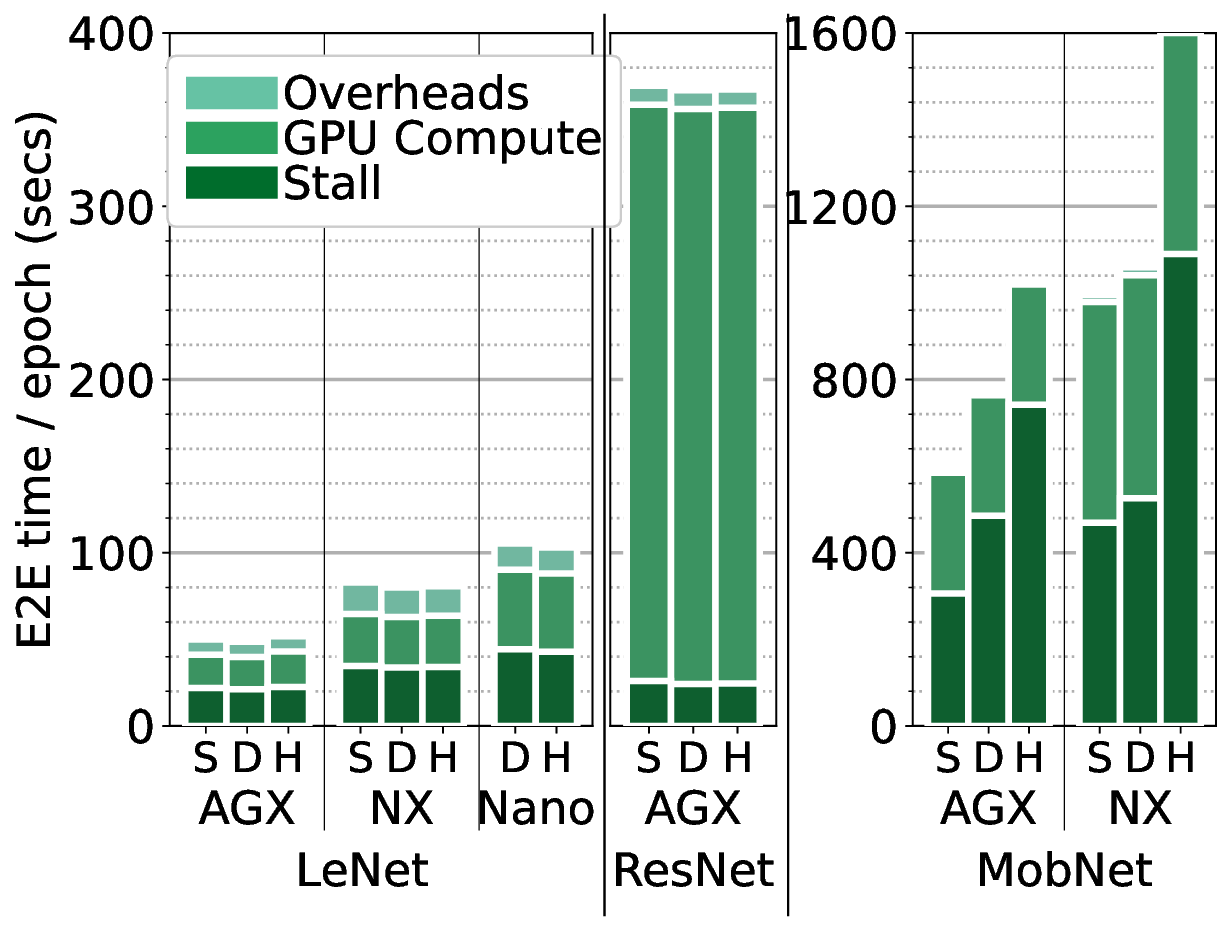}
    \label{fig:disk:stackede2e:e0}
    }\\
  \subfloat[Epochs $1+$ with $w=4$ ]{
    \includegraphics[width=0.49\textwidth]{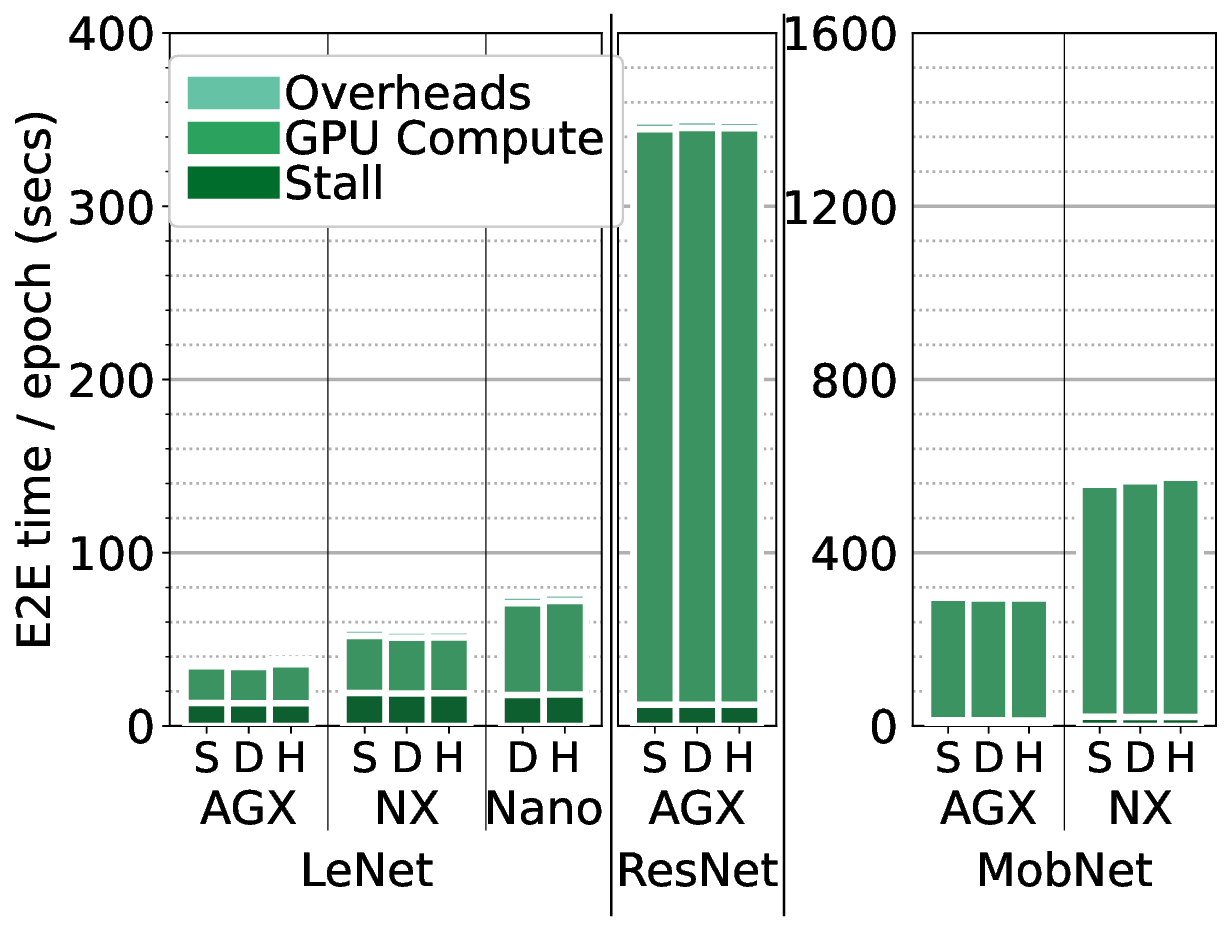}
    \label{fig:disk:stackede2e:e1+}
  }~~
  \subfloat[Energy for Epochs $1+$ with $w=4$]{
    \includegraphics[width=0.49\textwidth]{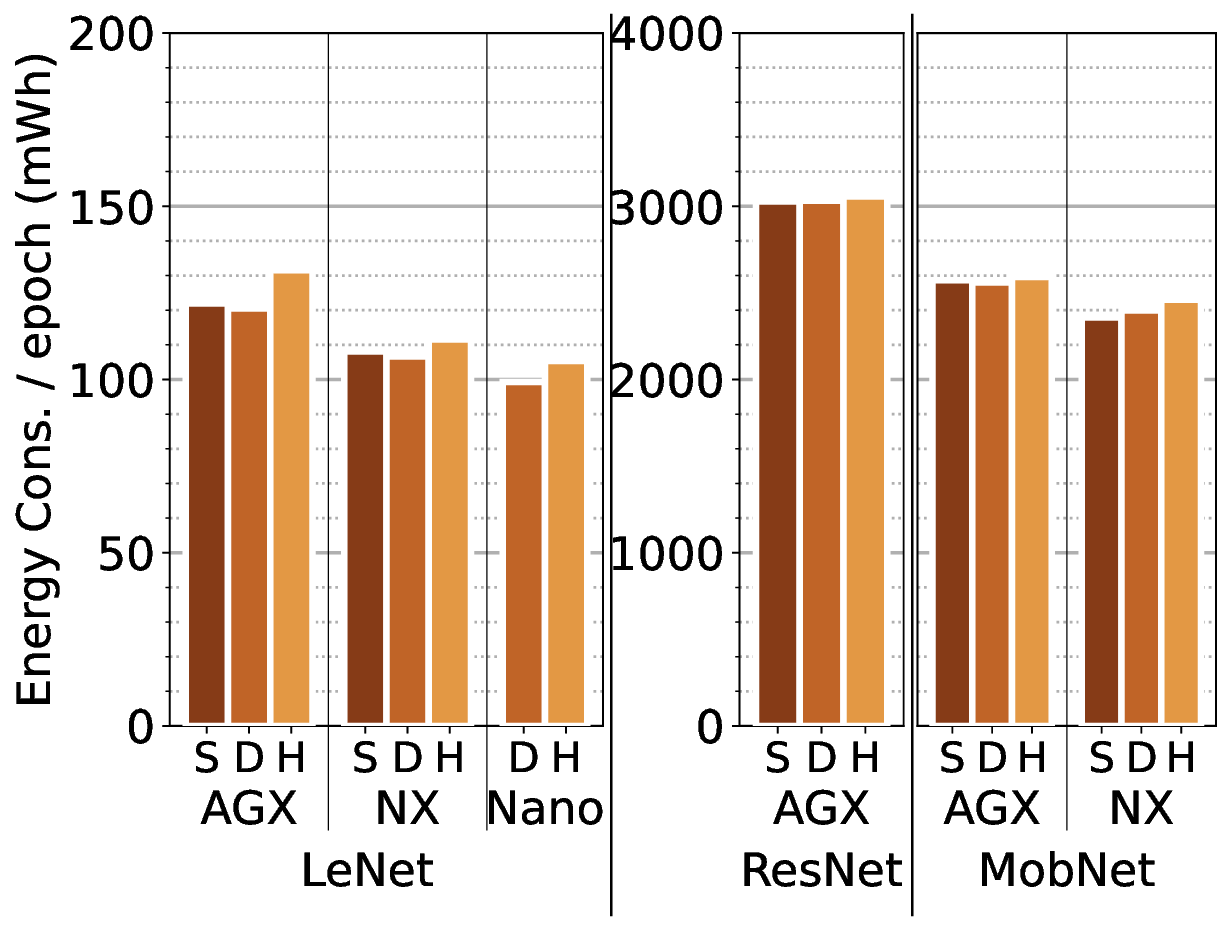}
    \label{fig:disk:energy:e1+}
  }
\caption{Effect of \underline{S}SD, S\underline{D} card and \underline{H}DD \textit{storage media} on the \textit{stall time} and the \textit{end-to-end time per epoch}.}
\label{fig:disk:stackede2e}
\end{figure*}

\vspace{-1ex}
\subsection{Effect of Storage Media}
\label{sec:analysis:disk}

Since the edge devices support a variety of storage media, it helps to understand the impact of these on the training time. This will allow us to select the appropriate storage type for a given training workload -- a faster (and costlier) disk may not necessarily give a performance benefit in certain cases. Here, we train the models on the devices using the default setup, but perform different runs with the training data present on S\underline{D} Card, \underline{H}DD, or \underline{S}SD, with the latter only supported on AGX and NX. Our observations and analysis are given below.

\claim{Any drop in stall time due to a faster storage media depends on the I/O pressure during fetch}

Fig.~\ref{fig:disk:stackede2e:e0} shows the stacked E2E time for epoch $0$ (i.e., no cache benefits) and with pipelining disabled ($w=0$) to localize the impact of disk speeds. When the mini-batch size for a model is small, such as MNIST and CIFAR, the I/O overheads of fetch are small. The IOPS are near zero in all cases.
As a result, having a faster SSD or SD card gives no stall time benefits for \lenet and \resnet. However, \mobilenet trains on the GLD data which loads $\approx 1.6MB$ from disk per mini-batch. This puts a higher I/O pressure on the disk and the difference between the three storage devices is visible. The stall time reductions match the disk speeds, 
with AGX reporting stall times of $306s$, $485s$ and $741s$ for SSD, SD card and HDD, respectively. 

\claim{Caching and pipelining can hide the stall times of a slower storage media, and a faster disk may not offer benefits}

For an expected training configuration of epoch $1+$ using $w=4$ pipelined and parallelized workers, the benefit of a faster disk is minimal.
Fig.~\ref{fig:disk:stackede2e:e1+} shows that the time taken is almost the same across disk media, for a given DNN model trained on a device. The stall times are small and their differences negligible, e.g., \mobilenet on AGX has stall times on SSD, SD card and HDD of $13s$, $13.1s$ and $12.9s$, relative to an overall E2E training time of $\approx 298s$ for all three.
While \mobilenet on NX is slightly slower for HDD, this is $<3\%$ relative to the SSD.

As a side-note, Fig~\ref{fig:disk:energy:e1+} shows that the storage media does not directly affect the energy per epoch, but the energy used changes due to the difference in training times.

\vspace{-1ex}
\subsection{Effect of Mini-batch Size}
\label{sec:analysis:bs}

\begin{figure*}[t]
\vspace{-0.15in}
\centering
  \subfloat[Stall Time]{
    \includegraphics[width=0.5\textwidth]{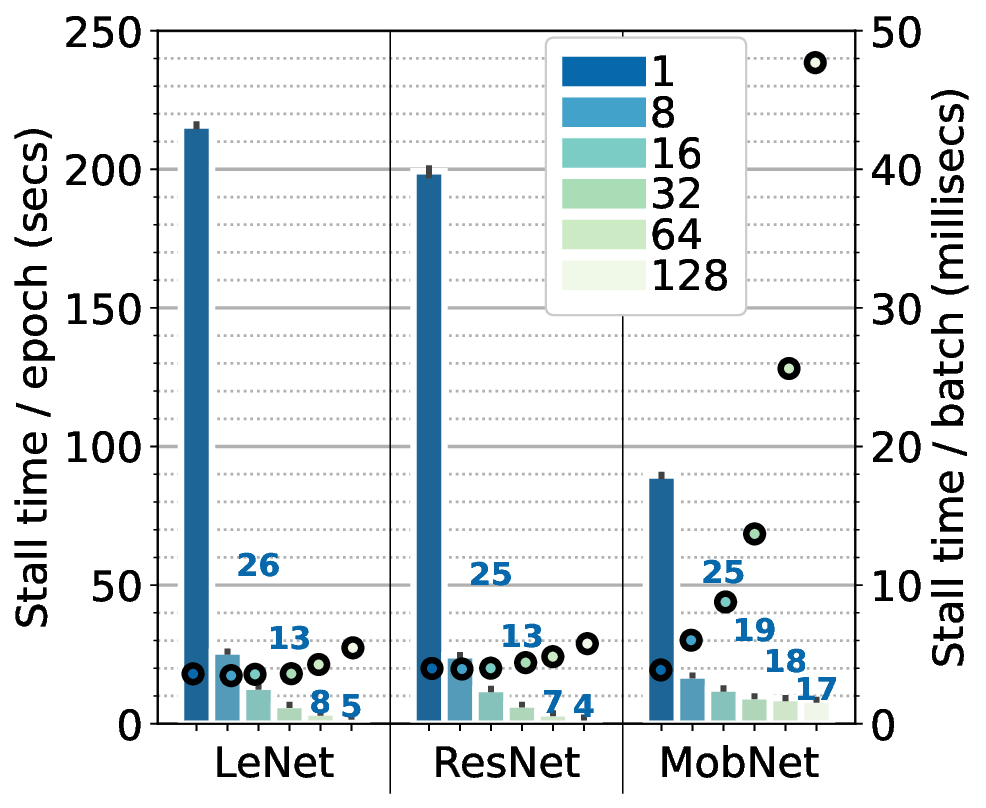}
    \label{fig:bs:stall}
  }\\
  \subfloat[GPU Compute Time]{
    \includegraphics[width=0.49\textwidth]{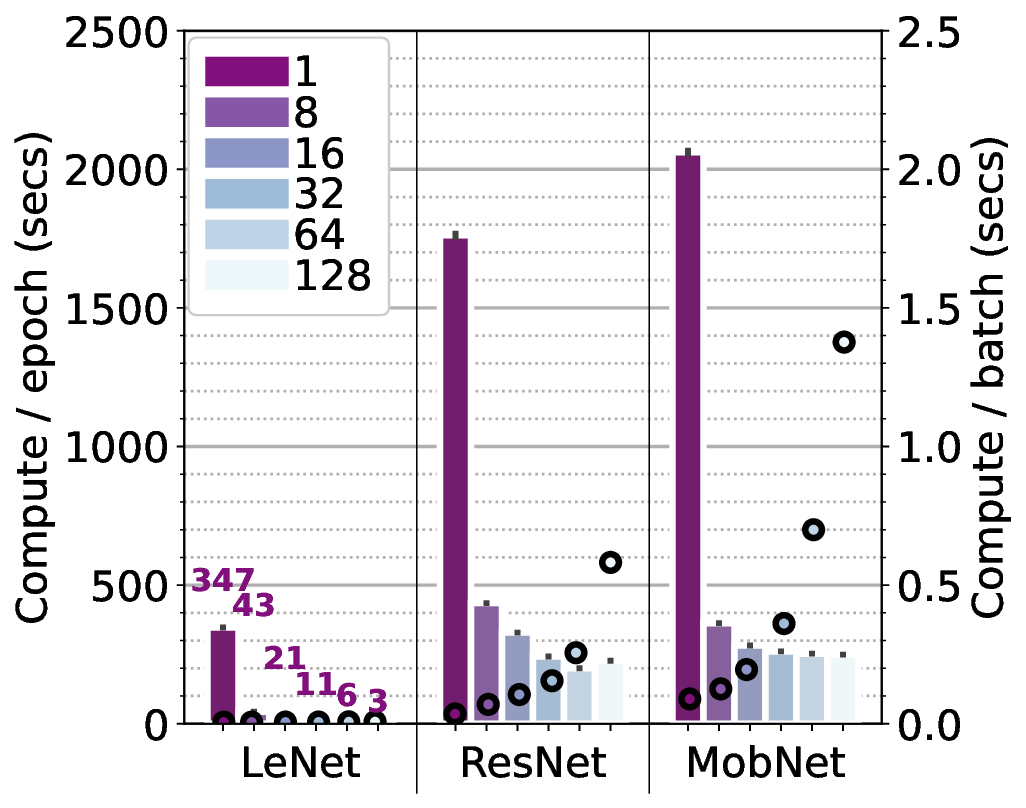}
    \label{fig:bs:gpu}
  }~~
  \subfloat[E2E time]{
    \includegraphics[width=0.49\textwidth]{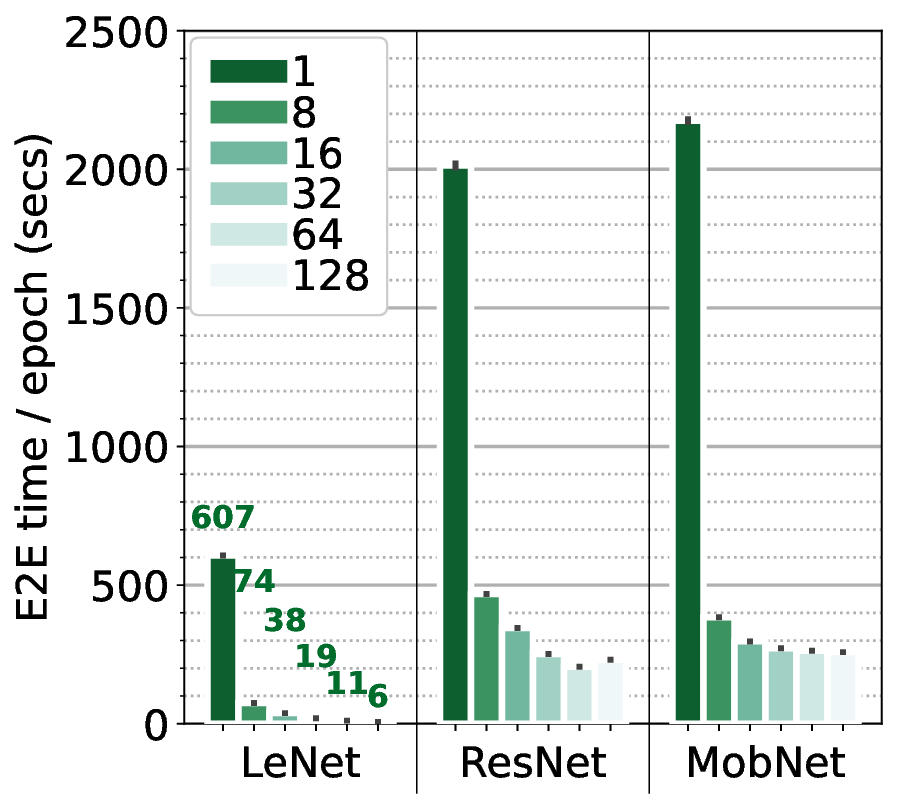}
    \label{fig:bs:e2e}
  }
\caption{Performance on AGX with \textit{batch-size ($bs$)} changing from $1$ to $128$. The \textit{time per epoch} is on the bars on the left Y axis, while the \textit{time per mini-batch} is on the markers on the right Y axis.}
\label{fig:bs}
\end{figure*}

Mini-batch size is a well studied hyper-parameter in DNN training, and it affects the statistical efficiency and rate of convergence. Their sizes range from $1$--$100s$ of samples~\cite{bengio_bs}, though smaller sizes of $2$--$32$ give better results~\cite{small_bs}. The maximum mini-batch size is limited by the GPU memory. Workstation GPUs like RTX2060 and RTX3080 with $6$--$12~GB$ of dedicated GDDR RAM can  run out of memory for larger models and mini-batch sizes. But the AGX has $32GB$ of RAM shared between CPU and GPU, and hence can train larger models and mini-batch sizes. So, it is worth examining the effect of varying the mini-batch size on the system performance. Here, we focus on AGX, which is the more recent edge device and has a larger memory.
We vary the mini-batch size from $bs=1$ to $128$ samples, but otherwise retain the default configuration for AGX from Sec.~\ref{sec:setup:default}.

\claim{Increasing the mini-batch size reduces the training time per epoch until the parallelism of the GPU cores saturate.}

As the mini-batch size increases, the data-parallelism of the GPU is better exploited as the samples in the mini-batch are independently processed on the different cores, and there are more rounds of data-parallel work per mini-batch. As a result, the compute time per mini-batch only gradually increases with larger batches until the GPU hits maximum utilization. 

We see this in Fig.~\ref{fig:bs:gpu} (markers on right Y axis), where for \mobilenet, increasing mini-batch size from $1$ to $8$ to $16$ increases the GPU compute time per batch by only $1.4\times$ and $1.56\times$. At $bs=16$ the GPU utilization is $99\%$ and doubling the mini-batch size almost doubles the time per mini-batch, e.g., from $195\mu s$ to $361\mu s$ from $bs=16$ to $bs=32$. With a larger mini-batch size, we have fewer batches per epoch and hence the total GPU compute time per epoch reduces. This is seen in the left Y axis bars where the compute time reduces sharply for all three models. The benefits plateau out for the larger models having $100\%$ GPU usage, e.g., beyond $bs=16$ for \resnet and \mobilenet, while they continue for the small \lenet model which uses only $25\%$ GPU even at $bs=128$.

\claim{Increasing the mini-batch size increases the stall time per mini-batch but reduces the overall stall time per epoch.}

As seen in Fig.~\ref{fig:bs:stall}, when the mini-batch size increases, there are more samples to be fetched per mini-batch. This involves more I/O, and also increases the CPU pre-processing time. This can be seen in the stall time per mini-batch increase on the right Y axis markers. Also, the GPU compute time per mini-batch grows slowly, as discussed above. This causes the stall time per mini-batch to increase. However, since the number of mini-batches per epoch decreases, the total stall time per epoch decreases, as seen in the bars on the left Y axis.

A bulk ($>90\%$) of the E2E time is a combination of stall time and GPU compute times. The stall time dominates for \lenet while the GPU compute time dominates for \resnet and \mobilenet. Fig.~\ref{fig:bs:e2e} shows the effects on the E2E time per epoch reducing due to both these factors, as $bs$ increases.

\subsection{Variability across Device Instances and Epochs}
\label{sec:analysis:var}

\begin{figure*}[t]
 \vspace{-0.15in}
\centering
    \subfloat[E2E Time]{
    \includegraphics[width=0.78\columnwidth,valign=b]{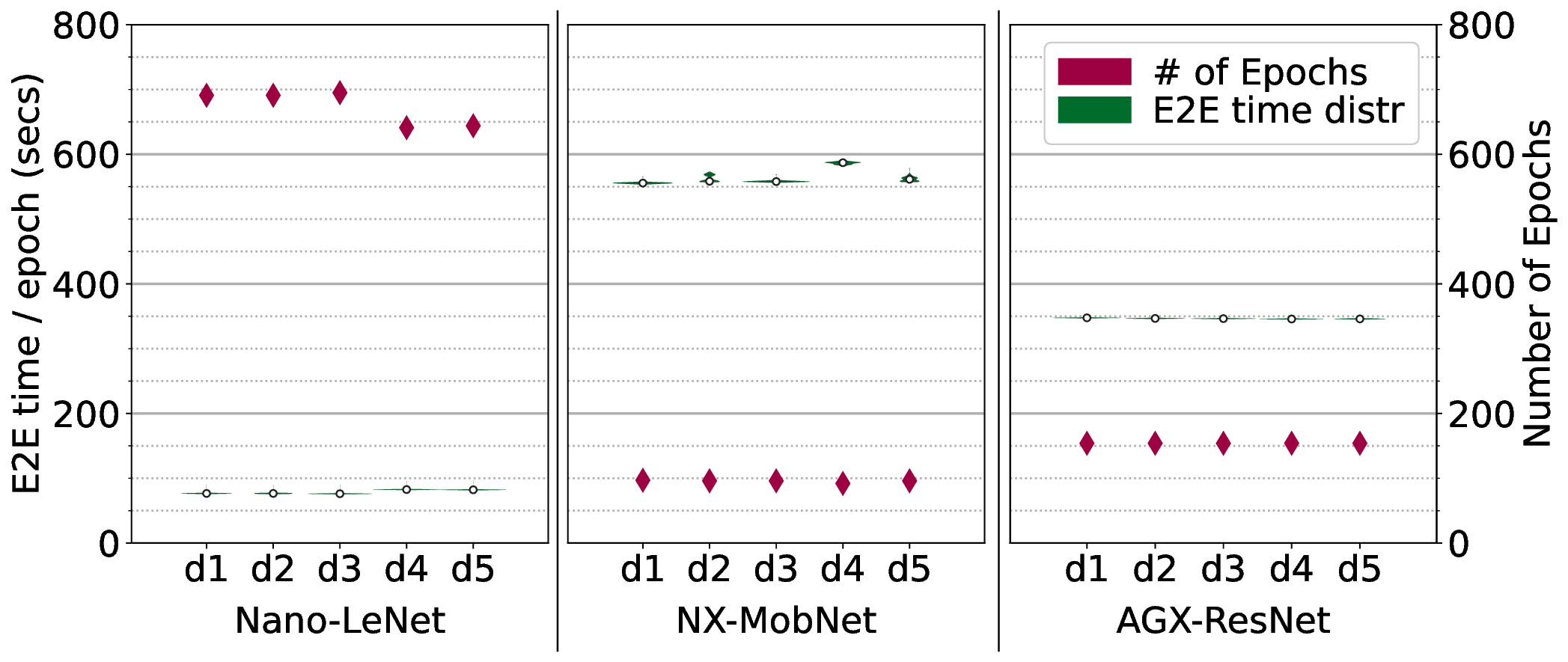}
    \label{fig:var:e2e}
  }\\
  \subfloat[Energy Consumed]{
    \includegraphics[width=0.78\columnwidth,valign=b]{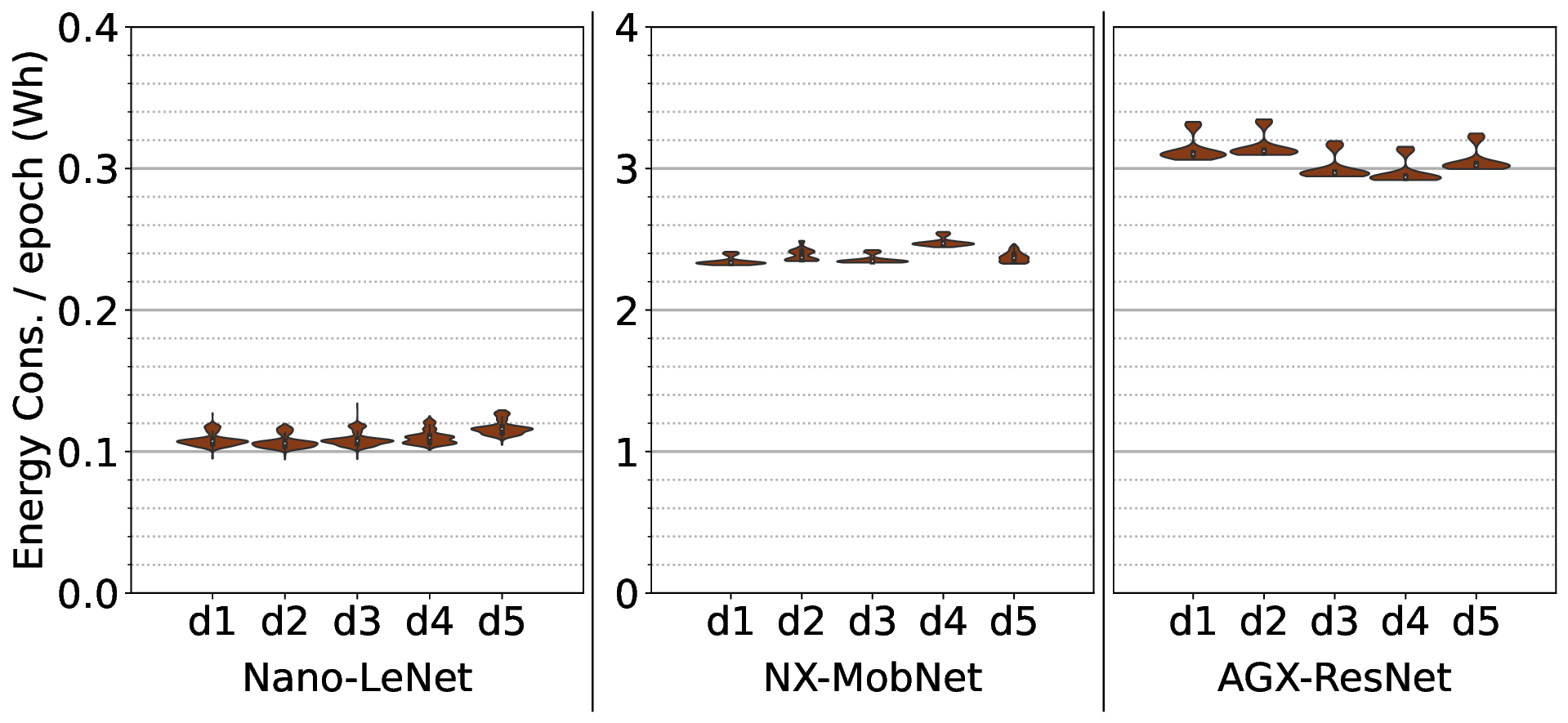}
    \label{fig:var:energy}
  }
\caption{Variability of device types and models for device instances over time for a $15h$ training run. Violin distribution of E2E time or Energy per epoch is on left Y axis. A marker for \# of epochs run is on right Y axis.
}


\label{fig:var}
\end{figure*}


A prior work on DNN inferencing on edge accelerators~\cite{snowflakes} reports a significant variability in both the inference latency and the power drawn for different instances of the same Jetson device type. In contrast to inferencing workloads that run in milliseconds, training workloads run for minutes or hours and are likely to be less sensitive across devices. Since DNN training can be long-running, we also study if there are changes in the device performance over a long training lifetime.

We train DNN models on $5$ devices each of AGX, NX and Nano. 
For each device type, we run the largest model it can train -- \resnet on AGX, \mobilenet on NX and \lenet on Nano. We run the training epochs continuously for a $15h$ period using the default configuration in Sec.~\ref{sec:setup:default}~\footnote{We conducted a $24h$ run on all the devices. However, we observe a sudden drop in power usage after $16h$ but with no impact on the training time per epoch or the resource performance. This is consistent across device types and instances, and across runs. We suspect it is a software overflow bug in the NVIDIA power monitoring tool, and are investigating this at the time of writing. Hence, we only report data for the first $15h$.}.

\claim{There is minimal variability in the E2E training time per epoch, for different epochs trained on a given device type}

Fig.~\ref{fig:var:e2e} shows a violin plot distribution of the E2E training time per epoch, for every instance of each device type. 
We see that all the violins have a tight distribution and the deviation across time even in the worst case is within $1\%$ for $\approx 150$ epochs of \resnet on AGX, $5\%$ for $\approx 95$ epochs of \mobilenet on NX and $1\%$ for $\approx 670$ epochs of \lenet on Nano.
So training a DNN for just a few epochs will generalize to more number of epochs on a device instance.

\claim{There is minimal variability in the E2E training time per epoch, across devices of the same type}

In Fig.~\ref{fig:var:e2e}, the median  E2E epoch training time across instances of a device type are almost identical. While they fall within $1\%$ for AGX, the variability is slightly higher for Nano and NX, at $7.8\%$ and $5.4\%$, due to marginal under-performance of Nano devices $d4$ and $d5$ and NX device $d4$.
So training a DNN on a single device will reasonably generalize to other instances of that device type.

\claim{There is minimal variability in the energy consumed per epoch, across time and across devices of the same type}

Fig.~\ref{fig:var:energy} shows that the energy consumed per epoch does not vary much across device instances or over different epochs. They fall to within $5.89\%$ for AGX, $5.53\%$ for NX and $8.81\%$ for Nano. These variations can be attributed to the minor changes in the training time per epoch. Issues like overheating, thermal throttling, etc. are not observed.

\claim{Significant variability is observed when there is a difference in the software configurations of devices}
Anecdotally, we observe that any changes to the OS kernel, PyTorch or NVIDIA Jetpack versions lead to variability in the performance of different instances of the same device type. In some cases, we see a variability of up to $40\%$ in the E2E time per epoch.
So careful attention has to be paid to the software setup across the devices to ensure reproducibility and deterministic behavior. E.g., we reboot each device before starting experiment runs to ensure a clean initialization.

\subsection{Effect of DVFS}
\label{sec:analysis:dvfs}

\begin{figure*}[t]
 \vspace{-0.15in}
\centering
  \subfloat[E2E time/epoch]
  {
    \includegraphics[width=0.35\columnwidth]{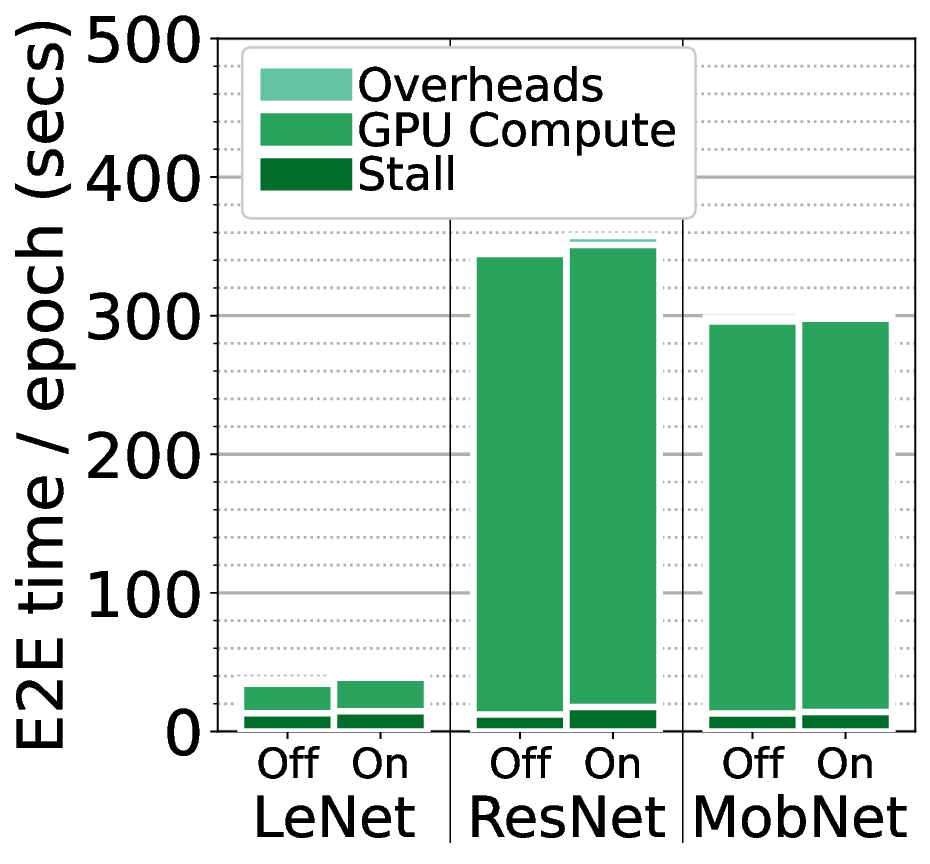}
    \label{fig:dvfs:e2e}
  }\quad
  \subfloat[Energy used/epoch]{
  \includegraphics[width=0.35\columnwidth]{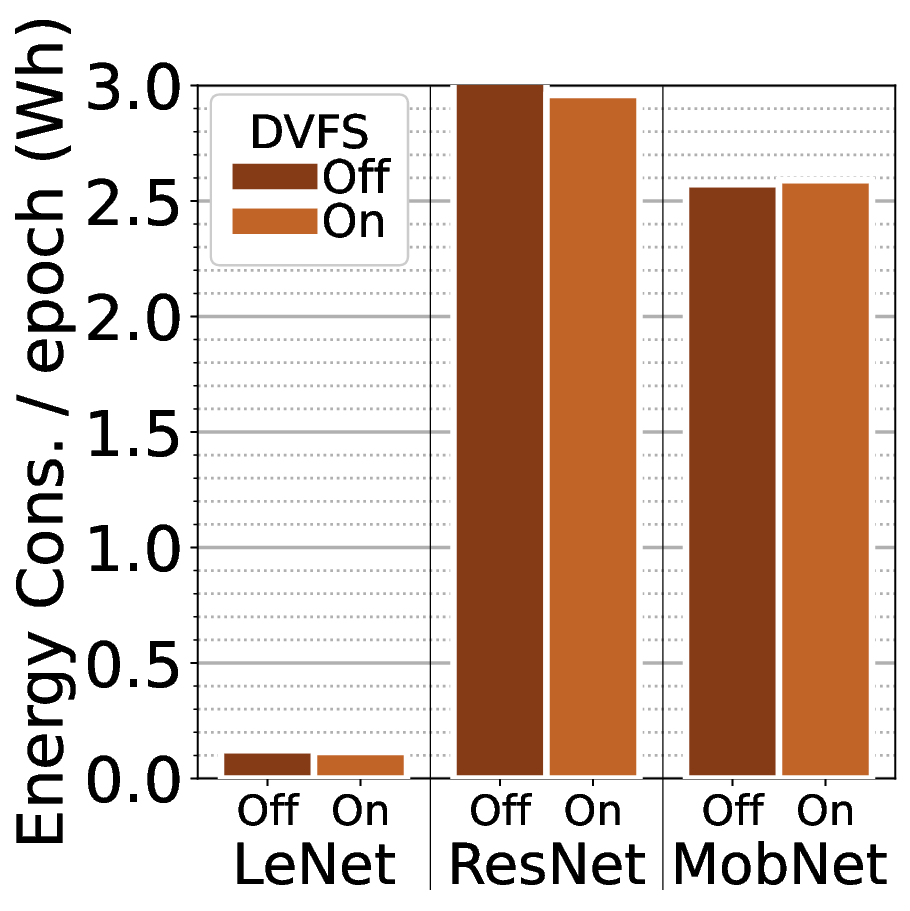}
    \label{fig:dvfs:energy}
  }
  
  \vspace{-0.125in}
  
  \subfloat[\%'ile of GPU Freqs.]{ 
    \includegraphics[width=0.35\columnwidth]{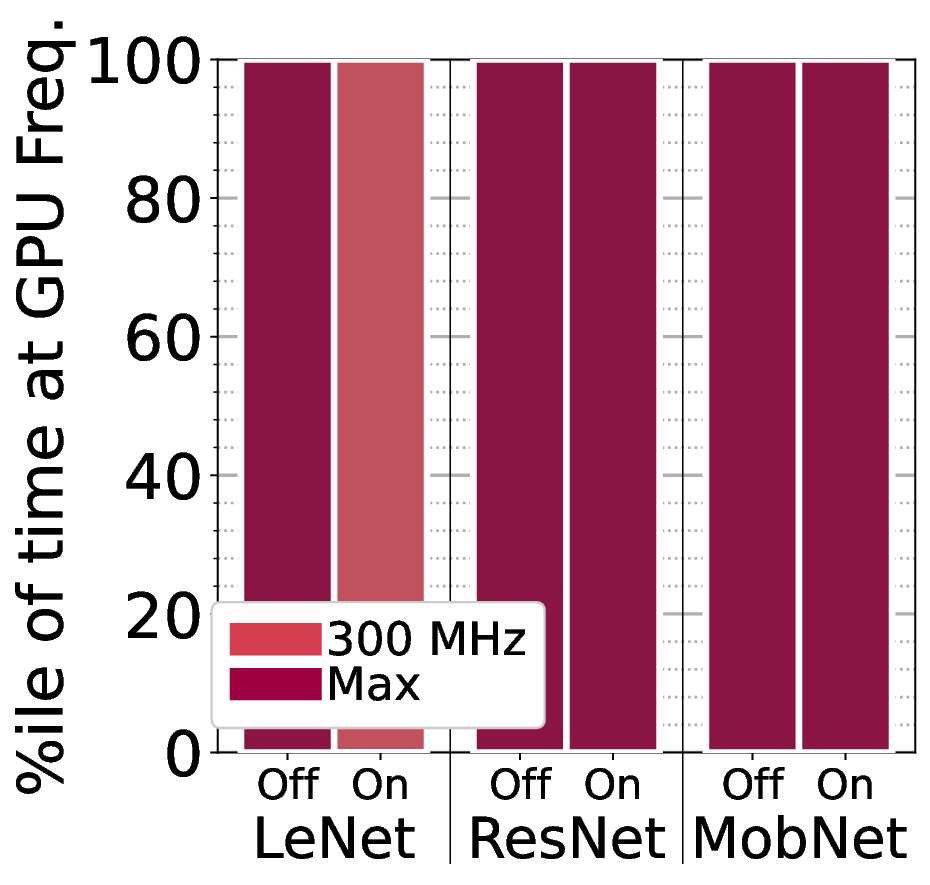}
    \label{fig:dvfs:gpu}
  } \quad
  \subfloat[\%'ile of CPU Freqs.
  ]{
    \includegraphics[width=0.35\columnwidth]{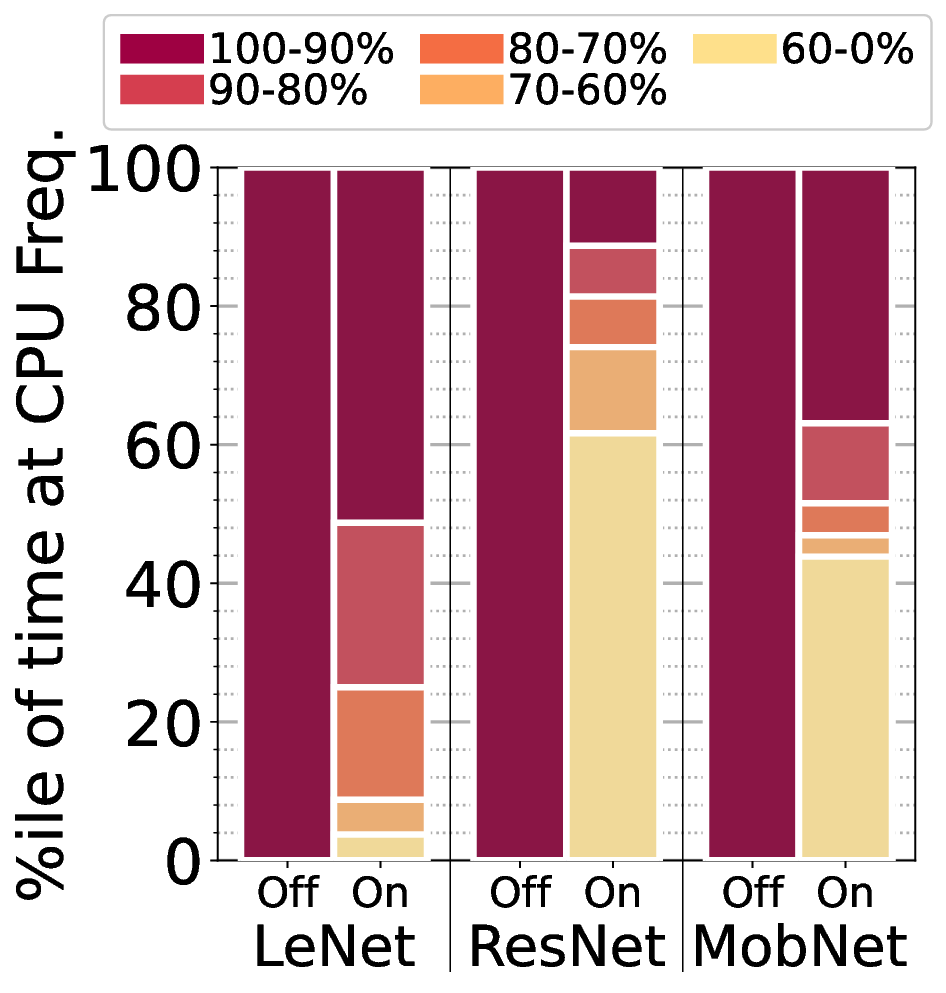}
    \label{fig:dvfs:cpu}
  }

\caption{Performance of AGX with DVFS off and on.}
 \vspace{-0.15in}
\label{agx_ssd_dvfs}
\end{figure*}

Enabling Dynamic Voltage and Frequency Scaling (DVFS) allows the CPU governor to dynamically alter the CPU, GPU and memory frequencies depending on the system load, to conserve energy. 
Here, we study the impact of DVFS on both the E2E time and the energy consumed. Here again, we limit our evaluation to AGX, for brevity. 
DVFS can be changed using the \texttt{jetson\_clocks} utility. We disable DVFS by setting the CPU, GPU and memory frequencies to a static value. We set this to the maximum frequency allowed for the default MAXN ($g$) power mode and do not change it. This ensures that the system always operates at peak performance~\footnote{In our experiments, we notice slight variations in frequencies even with DVFS off, e.g., when the CPU is set as $2265MHz$ we notice readings of $2263MHz$ and $2262MHz$. For simplicity, we consider all frequencies beyond the $95^{th}$ percentile of the expected value to be the static value.}. 
Other configurations are the defaults for AGX from Sec.~\ref{sec:setup:default}.

\claim{Enabling DVFS has negligible effect on the E2E time or the energy consumed per epoch.}

Fig.~\ref{fig:dvfs:e2e} shows the E2E time taken and the energy consumed per epoch, with DVFS off and on.
There is not much variation in either the time or energy for all $3$ DNN models. E.g., the most deviation we see is 
for \resnet with the E2E time being $349s$ and $358s$ with DVFS off and on, which is within $2\%$ of each other.
The energy usage difference is negligible as well with at most a $3\%$ variation seen, again for \resnet at $3.03Wh$ and $2.95Wh$.

\claim{Enabling DVFS does change the frequencies of CPU, GPU and memory, despite not affecting the training performance.}

While the values of E2E time and energy are similar with DVFS on or off, the frequencies are indeed changing when DVFS is on. E.g., \lenet has a low GPU utilization of about $7\%$ with DVFS off, and this causes the GPU frequency to reduce from $1377MHz$ to $300MHz$, in Fig.~\ref{fig:dvfs:gpu}, for the entire training period with DVFS on. This also causes the GPU utilization to increase to $14$--$34\%$, which helps it train within a similar E2E time. \resnet and \mobilenet have a high GPU utilization of $\approx 100\%$ and hence the GPU frequency is not modified.
The memory frequencies (not plotted) also show no differences for the latter two models, and some reduction in frequency for $16\%$ of the time for \lenet.

We see a more active variation in the CPU frequencies across the models.  Fig.~\ref{fig:dvfs:cpu} shows the fraction of training times where the CPU was set to different frequency values, given as a $\%$ of the maximum frequency, $2265MHz$. \lenet exhibits CPU variability for $\approx 50\%$ of its runtime, while it is at $>90\%$ of peak CPU clock speed for the rest. This is both due to the pre-processing costs and the higher overheads for launching the GPU kernels for each layer.
The CPU frequencies for \resnet and \mobilenet are below their peak clockspeed for $90\%$ and $65\%$ of their runtime, respectively. Since they are GPU bound, the CPU has low utilization for the most part. E.g., with DVFS off, the median CPU utilization for \resnet and \mobilenet are only $11\%$ and $25\%$.

\subsection{Baseload and Effect of Power Modes}
\label{sec:analysis:power}

\textbf{Baseload under Idle States.} In this section, we drill-down into some of the power modes of the Jetson devices. Prior to that, it is helpful to understand the baseload on the devices under different states of idleness, to understand their sustainability and energy impact. We examine four baseload conditions: (i) A clean-start of the device with no applications running, but with the logging of performance and energy metrics turned on and DVFS turned off. This setting mimics our default experiment harness but without actually running any training workloads. (ii) No applications or logging harness running, but DVFS turned off. (iii) No applications or logging running, but DVFS turned on. (iv) Device in Wake on LAN (WoL) state, where it can be activated by a network command but is otherwise suspended. The last is helpful when devices have to be occasionally woken up for training but can otherwise remain turned off.

\begin{figure}[ht]
    \centering
    \includegraphics[width=0.5\columnwidth]{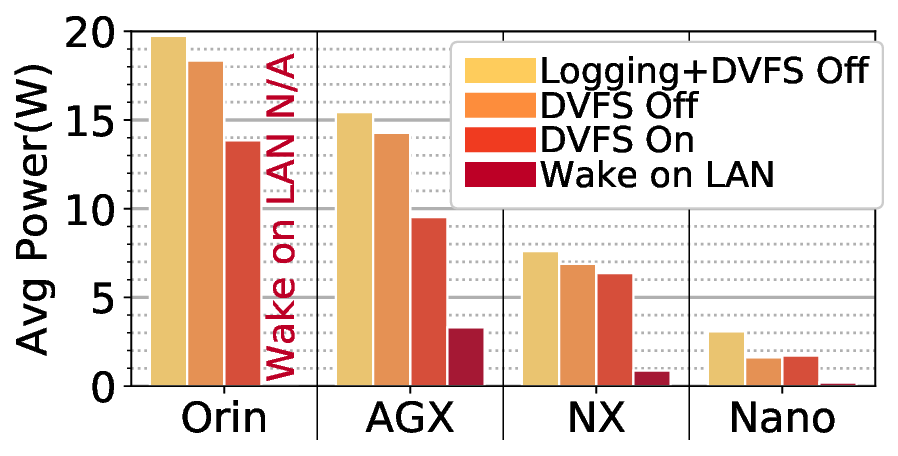}
    \caption{\textit{Average socket power load (W)} for various idle states on all devices.}
    \label{fig:baseload}
\end{figure}

Since logging is disabled in (ii)--(iv), we instead use the JouleJotter~\cite{JJ} power monitor in these baseload experiments. It samples the plug-load to the device every $20s$ and records it locally.
All devices are set to their default (MAXN or equivalent) power modes, and we measure their power loads over a $30~min$ period.
The average of these samples for the devices and idle states is reported in Fig.~\ref{fig:baseload}. Orin's WoL was not supported as of the time of writing. 

As expected, WoL has the lowest power consumption of all the idle states for all devices, and is substantially lower than the next lowest idle state with DVFS on. On the Nano, WoL uses $<1W$ of power while it is $<4W$ even for AGX. Turning on DVFS results in a significant power saving for more powerful devices like Orin and AGX, but has negligible impact on NX or Nano. Logging has minimal impact on the faster devices, while it leads to higher power load for Nano. This is a constant overhead during the training experiments.
The power load during training tends to be much higher since the CPU and/or GPU are active. E.g., the average power load on AGX when training \resnet is $31W$ while it is $5W$ when training \lenet on Nano.



\begin{figure}[ht]
    \centering
    \includegraphics[width=0.5\columnwidth]{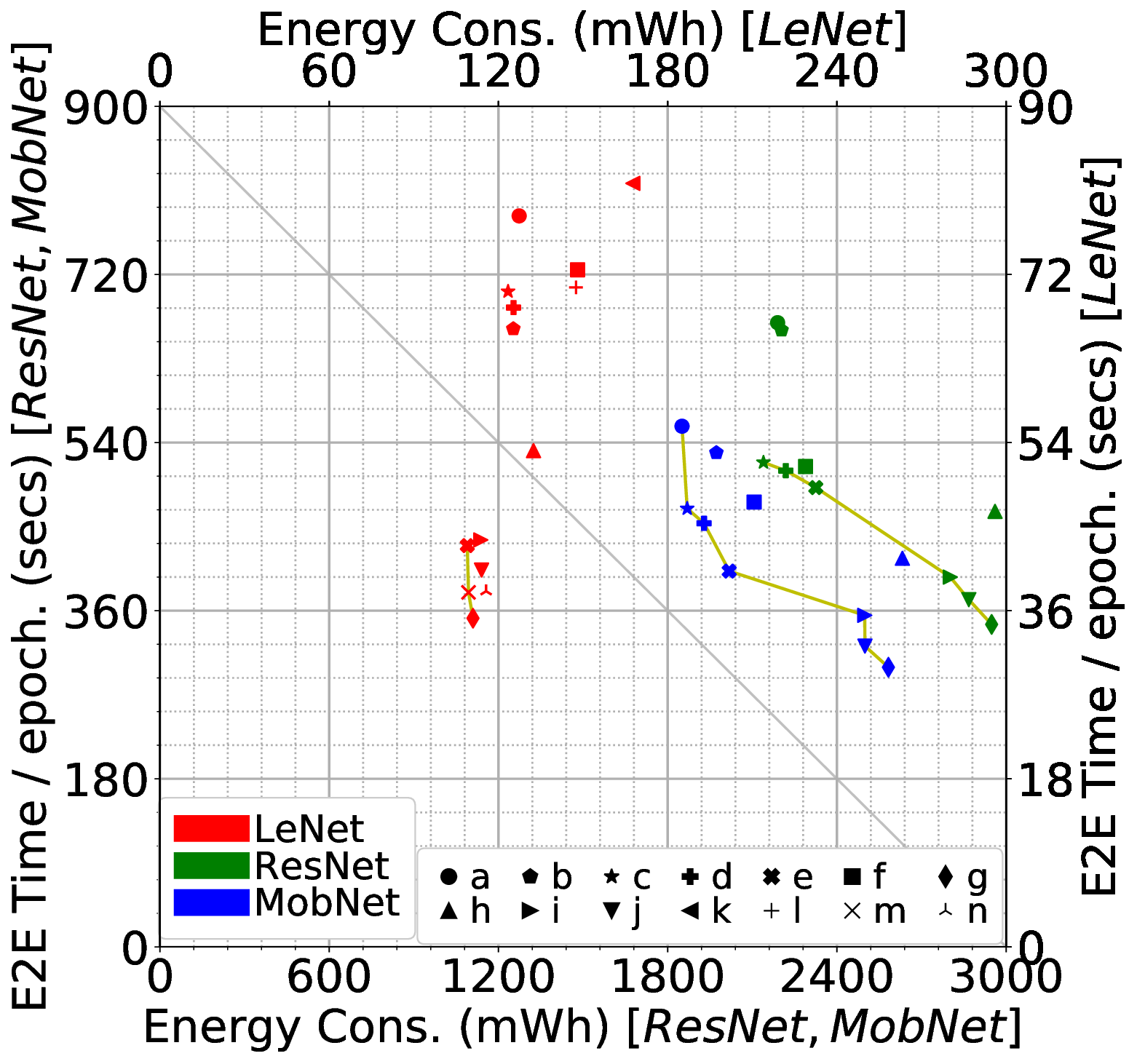}
    \caption{Scatter plot of \textit{E2E time vs. Energy consumed per epoch ($1+$)}, for power modes $a$--$n$ of AGX for \lenet (secondary axes) and \resnet/\mobilenet (primary axes). The yellow lines indicate the \textit{Pareto front} per DNN.}
    \label{fig:power:scatter}
\end{figure}

\begin{figure}[ht]
    \centering
    \subfloat[E2E time]{
        \includegraphics[width=0.85\columnwidth]{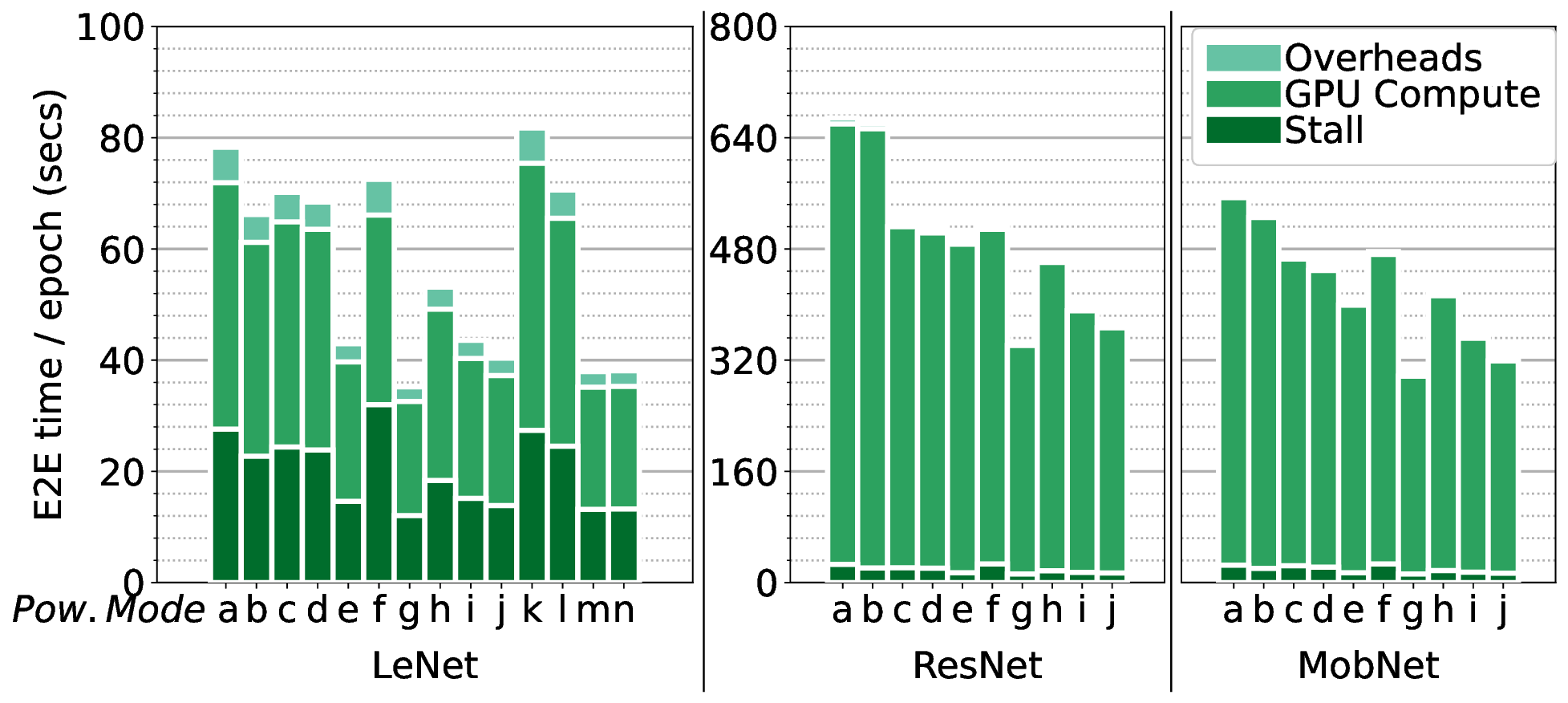}
        \label{fig:power:e2e:pm}
    }\\
    \subfloat[Energy]{
        \includegraphics[width=0.85\columnwidth]{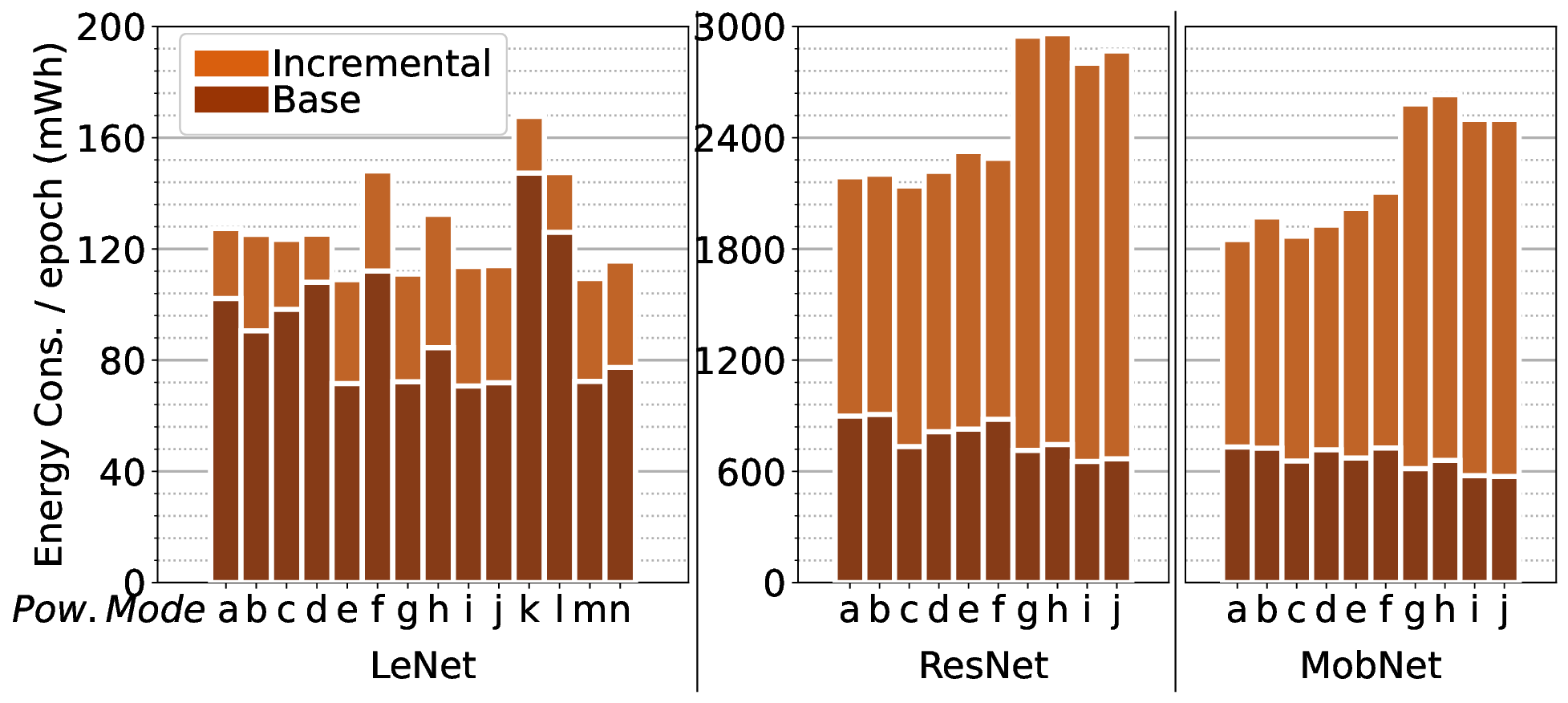}
        \label{fig:power:energy:pm}
    }
    \caption{E2E time and energy usage per epoch for different power modes of AGX. See Table~\ref{tbl:power} for power mode labels.}
    \label{fig:power}
\end{figure}


\textbf{Impact of Power Modes.} The Jetson devices come with a number of pre-defined power modes that users can choose from. Additionally, we can also configure a custom power mode by specifying the number of CPU cores enabled, and the frequencies of CPU, GPU and RAM (External Memory Controller (EMC))~\footnote{Not all frequencies are allowed. There are $29$ possible CPU frequencies, $14$ GPU frequencies and $9$ memory frequencies.}
The power mode can be \textit{changed on the fly}, without any system downtime. 

This can help define an ideal power mode for each DNN model training, which balances the training time and the energy used by the constrained edge device, e.g., to stay within a daily energy budget or to avoid overheating of enclosures, while still minimizing the training time.
The range of values for the frequencies is also wide, and choosing one power mode over another can result in an order of magnitude performance difference in time and energy. E.g., running Resnet on AGX using the MAXN peak power mode is $10.3\times$ faster, has $3.6\times$ more power load and consumes only $0.4\times$ energy compared to running it with a much lower GPU frequency of $114.75MHz$.

We study the impact of power modes on the training time and energy use of AGX.
We choose a mix of both pre-defined and custom power modes, labelled $a$--$n$ in Table~\ref{tbl:power}. We typically vary one resource parameter at a time (shown in bold) between the modes to examine its incremental impact. Some power modes are only evaluated for specific experiments.
We use the defaults for AGX (Sec.~\ref{sec:setup:default}) and report results for epochs $1+$.

\begin{table}[t]
\centering
\def\arraystretch{0.5}
\footnotesize
\vspace{-0.1in}
\caption{Power Modes Evaluated}
\vspace{-0.1in}
\label{tbl:power}
\begin{tabular}{l|c| rrrr}
\toprule
 \bf{Device} & \bf{Label} & \bf{CPU Cores} & \bf{CPU MHz} & \bf{GPU MHz} & \bf{RAM MHz} \\
\hline
 \multirow{14}{*}{AGX} & \textbf{a}  & 4 & 1200 & 670 & 1333 \\
   \cmidrule{2-6}
  & \textbf{b}  & \textbf{8} & 1200 & 670 & 1333\\
 \cmidrule{2-6}
  & \textbf{c}  & 8 & 1200 & \textbf{900} & 1333 \\
 \cmidrule{2-6}
  & \textbf{d}  & 8 & 1200 & 900 & \textbf{1600} \\
 \cmidrule{2-6}
  & \textbf{e}  & 8 & \textbf{2100} & 900 & 1600\\
 \cmidrule{2-6}
  & \textbf{f}  & \textbf{2} & 2100 & 900 & 1600\\
 \cmidrule[1pt]{2-6}
   & \textbf{g (MAXN)} &  \textbf{8} & \textbf{2265} & \textbf{1377} & \textbf{2133}\\
  \cmidrule{2-6}
  & \textbf{h} & 8 & 2265 & 1377 & \textbf{1066} \\ 
 \cmidrule{2-6}
  & \textbf{i}  & 8 & 2265 & 1377 & \textbf{1333} \\ 
 \cmidrule{2-6}
  & \textbf{j}  &  8 & 2265 & 1377 & \textbf{1600}\\ 
 \cmidrule[1pt]{2-6} 
  & \textbf{k}  & \textbf{4} & \textbf{1036} & \textbf{420} & \textbf{2133}\\
 \cmidrule{2-6}
  & \textbf{l}  & \textbf{8} & 1036 & 420 & 2133\\
 \cmidrule{2-6}
  & \textbf{m}  & 8 & \textbf{2265} & 420 & 2133\\
 \cmidrule{2-6}
  & \textbf{n}  & 8 & 2265 & \textbf{900} & 2133\\
\hline
 NX & \textbf{15W} & 6 & 1400 & 1100 & 1600 \\
\hline
Nano & \textbf{MAXN} & 4 & 1479 & 921 & 1600 \\
\midrule
Orin & \textbf{MAXN} & 12 & 2200 & 1300 & 3200 \\
\bottomrule 
\multicolumn{6}{c}{\textit{Cells in bold indicate a value change from the cell in the previous row.}}\\
\multicolumn{6}{c}{\textit{Since most of these are custom power modes, they do not have a preset power budget.}}
\end{tabular}
\vspace{-0.2in}
\end{table}

In Fig.~\ref{fig:power:scatter}, we plot the E2E time (Y axis) and the energy consumed per epoch (X axis) for the $3$ DNN models and $14$ power modes evaluated. We draw the Pareto front (yellow lines) for each device, which is the envelope that minimizes both the epoch training time and the energy, and is monotonic along one or both axes. We first add the leftmost data point (lowest energy) to the Pareto front. As we move right (increasing energy), any data point whose E2E time is lesser than the previous data point is added to the Pareto front. These Pareto optimal points have the least Y-axis value for any X-axis value, or vice versa.
While \resnet and \mobilenet offer several Pareto optimal points for an optimization trade-off, \lenet has just three, limiting the choices. 
We also show variations of this scatter plot to highlight the effect of each of resource type in the Appendix.
A similar plot of E2E time against the average power load (Fig.~\ref{fig:power:time:scatter} in the Appendix) shows an inverse correlation, as expected, with the training time per epoch decreasing as the power load increases for a higher power mode. 

\claim{The default power mode may not be Pareto optimal.}
For all models, the peak power mode $g$ (MAXN; $\blacklozenge$) with the highest core counts and frequencies has the fastest time, and is on Pareto front.
But the system default power mode for AGX is $a$. This mode is on the Pareto front for \mobilenet but not for \lenet or \resnet. So training with the default power mode may give a sub-optimal time--energy trade-off, depending on the model, necessitating an intelligent choice of power mode.

\claim{The energy consumed is often dominated by the baseload rather than the incremental load due to training.}

We record the energy consumed by the AGX under baseload, i.e., when the system is on and doing minimal processing like monitoring resource counters (idle state (i)). The baseload for $g$ (MAXN) is $7527~mW$. As Fig.~\ref{fig:power:energy:pm} shows, the baseload (dark brown) forms a large fraction of energy consumed when training, i.e., the device would have consumed the same energy for that period even if we were not training a model. This is larger at $65$--$80\%$
for smaller models like \lenet, which do not use much CPU and GPU, while it is about $20$-$45\%$ for \resnet and \mobilenet.
So, unless the device is in a sleep state with WoL activation only during training, it is better to put it to use for training rather than stay idle and consume similar energy.

\claim{The variation in energy consumed per epoch is modest across power modes for larger DNN models.}

While we experiment with a wide range of power modes, the energy consumed per epoch does not vary much across them. The reduction between the most and the least energy consuming mode (excluding MAXN and its neighbors, $g$--$j$), is $\approx 27\%$ for \lenet,
$\approx 6\%$ for \resnet and $\approx 12\%$ for \mobilenet, as seen in Fig.~\ref{fig:power:energy:pm}.
So, for the larger models, choosing a power mode that trains faster may not impose a higher energy penalty. MAXN is the only exception, and tends to consume more cumulative energy to complete training than others.

\claim{The GPU compute time is inversely proportional to the GPU frequency for larger DNNs.}

As expected, increasing the GPU frequency reduces the compute time spent on the GPU, but this is limited to larger DNN models that are GPU-bound.
In going from power mode $b$ to $c$, the GPU frequency increases from $670MHz$ to $900MHz$ while all other parameters stay the same. 
The drop in GPU compute time (Fig.~\ref{fig:power:e2e:pm}) for \resnet is $22.3\%$, from $632s$ to $491s$. \mobilenet has a more modest drop of $12.6\%$ from $505s$ to $441s$. \lenet is affected the least, in fact increasing from $38.4s$ to $40.5s$, as it is lightweight and does not utilize the GPU fully. 

\claim{Increasing the CPU frequency and number of cores reduce the stall time.}

The stall time depends on the difference between fetch and pre-processing time, and the GPU compute time. Increasing the CPU frequency and/or core count can reduce the pre-processing time and hence the stall time.
In going from power mode $d$ to $e$, the CPU frequency steeply rises from $1.2GHz$ to $2.1GHz$, causing a drop in stall time by $31$--$38.7\%$ for all three models. While stall time is a small part of the E2E time for \mobilenet and \resnet, this leads to a $13.5\%$ improvement in E2E time for \lenet. 

Similarly, when we double the core-count from $4$ to $8$ for power modes $a$ to $b$, we see a drop in stall time by $17.6$--$19.5\%$ for the three models. Here, the benefits are incremental since we have only $w=4$ worker processes and increasing beyond $4$ cores does not help much.
However, when we drop from $8$ to $2$ cores between power modes $e$ and $f$, the stall time sharply increases by $54.4\%$ for \lenet and $\approx 47\%$ for \mobilenet and \resnet. Since we have fewer cores than the active worker processes, the worker parallelism suffers.

\claim{Increasing the CPU frequency and number of cores reduces the GPU compute time, and more so for light models.}

Besides pre-processing, the CPU also loads the kernel to the GPU. Depending on the DL framework used, there may be one or more kernels launched per DNN layer. 
This time can be significant for lightweight DNNs~\cite{KL_mobile}. Hence changing the CPU frequency and core-count affects the GPU compute time, which includes the time for the kernel launch. 

In Fig.~\ref{fig:power:e2e:pm}, when going from power mode $d$ to $e$, the CPU frequency jumps from $1.2GHz$ to $2.1GHz$ while other resources stay the same. For \lenet, this causes a significant decrease in GPU compute time by $36.8\%$, while this is smaller at $9.7\%$ and $2\%$ for \mobilenet and \resnet.

Alternatively, when the CPU cores drop from $8$ to $2$ between power modes $e$ and $f$, \lenet sees an increase in GPU compute time by $35.9\%$. Since the kernel is launched multiple times per mini-batch, one of the CPU cores will be busy for this operation. There will be contention for cores between the kernel launch, the $4$ workers and the logging process when there are fewer than $6$ cores. 
\mobilenet is computationally costlier than \lenet, and can amortize the kernel launch overheads better. It shows a smaller increase of $15.6\%$. \resnet which is the most demanding computationally is the least affected, with a minor increase of $1.9\%$. The effect of fewer cores when pipelining and parallelism are disabled ($w=0$; not plotted) is less prominent, at $20.8\%$ for \lenet and negligible for \mobilenet and \resnet due to less contention.

We illustrate the effects of CPU on the GPU compute time through specific runs on \lenet for modes $k$--$n$. In moving from $l$ to $m$, as CPU frequency increases from $1036MHz$ to $2265MHz$, the GPU compute time drops by $46.3\%$ (Fig.~\ref{fig:power:e2e:pm}). Similarly, when the CPU cores increase from $4$ to $8$ for $k$ to $l$, the GPU time drops by $14.6\%$.
But, when the GPU frequency increases from $420MHz$ to $900MHz$ to $1377MHz$ between $m$ to $n$ to $g$, the GPU time does not improve by more than $6.8\%$. Here, GPU compute time is more sensitive to CPU frequency/cores than GPU frequency. Thus, running \lenet in a high CPU frequency mode gives the best E2E time, and the lowest energy (Fig.~\ref{fig:power:energy:pm}).

\claim{GPU compute time and stall times are affected by the memory frequency.}
To understand the effect of memory frequency, we evaluate power modes $g$--$j$ where only the memory frequency varies. 
From Fig.~\ref{fig:power:e2e:pm}, we see that as memory frequency increases, the GPU compute time decreases. E.g., when the frequency doubles from mode $h$ to $g$, the GPU compute time drops by $33.4\%$ for \lenet, $25.8\%$ for \resnet and $28\%$ for \mobilenet. We also see that the stall times reduce by $34.4\%$, $27.5\%$ and $28.1\%$ for these models. This indicates that the frequency of the memory shared between CPU and GPU has a tangible impact on the training performance.

\vspace{-1ex}
\subsection{Predicting Training Time and Energy Usage for Custom Power Modes}%
\label{sec:analysis:predict}

Besides the insights drawn from our experiments above, we can also use the empirical data to model the device behavior under training.
We present a feasibility study on predicting the time and energy required to train a \textit{candidate DNN model} for any custom power mode of a given device. We use simple regression techniques and minimal \textit{a priori} profiling for a limited number of epochs and power modes.
This can help select the best power mode to trade-off time and energy for model training, or for selecting a subset of (heterogeneous) devices for federated learning~\cite{google-sysml}.

Our results show that training our time prediction model for just $3$ epochs and for the \textit{$4$ power modes we recommend} -- which takes about $2~h$ even for the costliest DNN we evaluate --
helps predict the training time for any power mode within $\approx 12\%$ 
Mean Average Percentage Error (MAPE). 
The energy usage is dominated by the baseload, and estimating the model training time also helps predict the energy usage.
We offer preliminary results from a single pre-trained linear regression model to predict the energy consumption per epoch.
In the future, more sophisticated techniques can help further improve the modeling accuracy. For brevity, we limit this study to AGX. 

\vspace{-1ex}
\subsubsection{Model Training to Predict E2E Time per Epoch}
We fit a simple \textit{linear regression model} for a given candidate DNN to predict its \textit{training time per epoch} by collecting its training time for $4$ power modes we identify and for $3$ epochs each; we drop epoch $0$ due to its bootstrapping overheads and use the remaining $4 \times 2 =8$ samples. We fit the equation:~~$a \cdot x1 + b\cdot x2 + c\cdot x3 + d\cdot x4 + e = T_i$~~
over these $8$ samples that form $T_i$, the training time per epoch. The input feature vector: \textit{CPU frequency ($x1$), CPU cores ($x2$), GPU frequency ($x3$)} and \textit{memory frequency ($x4$)} is set by the power mode.

A key contribution is identifying the \textit{$4$ representative power modes} to train the candidate DNN over for a good prediction of the training time. We perform an exhaustive training of regression models from the $_{10}C_4$ possible combinations of $4$ power modes chosen from the $10$ modes that we evaluated for all 3 DNNs in Sec.~\ref{sec:analysis:power}. Among these, we select the set whose regression fit returns the lowest sum of Root Mean Square Error (RMSE) for training times predicted for the 3 DNNs: \lenet, \mobilenet and \resnet. The power modes thus identified are: $e$,$f$,$i$ and $j$, which cover 2 different core-counts, CPU and GPU frequencies, and 3 different memory frequencies.

So, for any new DNN model, the user runs $3$ epochs for the power modes $\{e,f,i,j\}$, takes their latter $2$ epoch training times to fit the first regression model, and uses this to predict the per-epoch training time for the DNN for any power mode.

For energy, we fit a common (universal) linear regression model over the training data collected for the various power modes for training $2$ epochs of the $3$ DNNs. It takes the power mode resource values and the (predicted) training time as input, and estimates the expected energy per epoch.

\begin{figure}[t]
 \centering
 \vspace{-0.15in}  
\subfloat[E2E Time and MAPE per power mode]{
    \includegraphics[width=0.7\columnwidth]{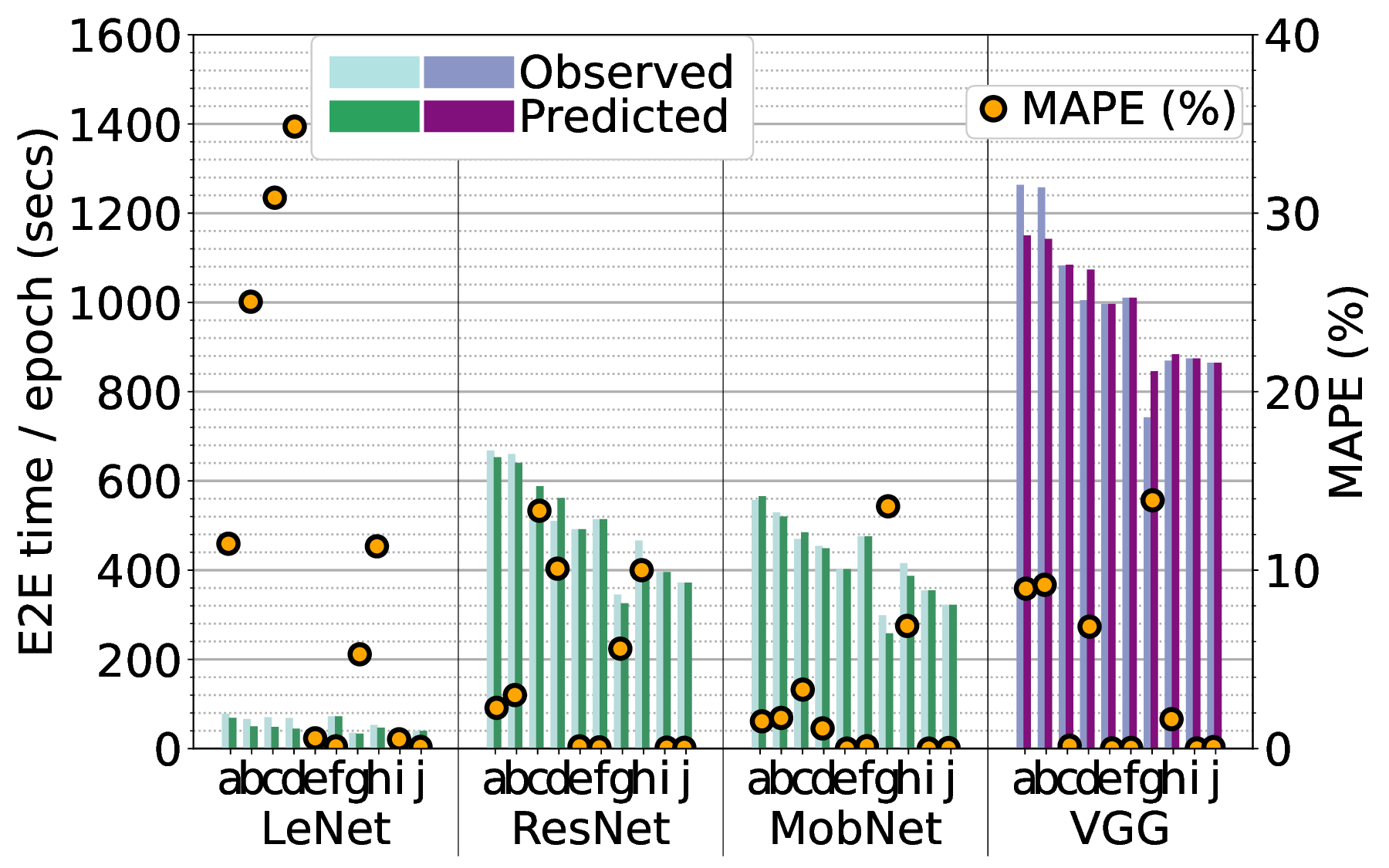}
    \label{fig:predict:e2e:mode}
  }\\
  \subfloat[Energy and MAPE per power mode]{
    \includegraphics[width=0.7\columnwidth]{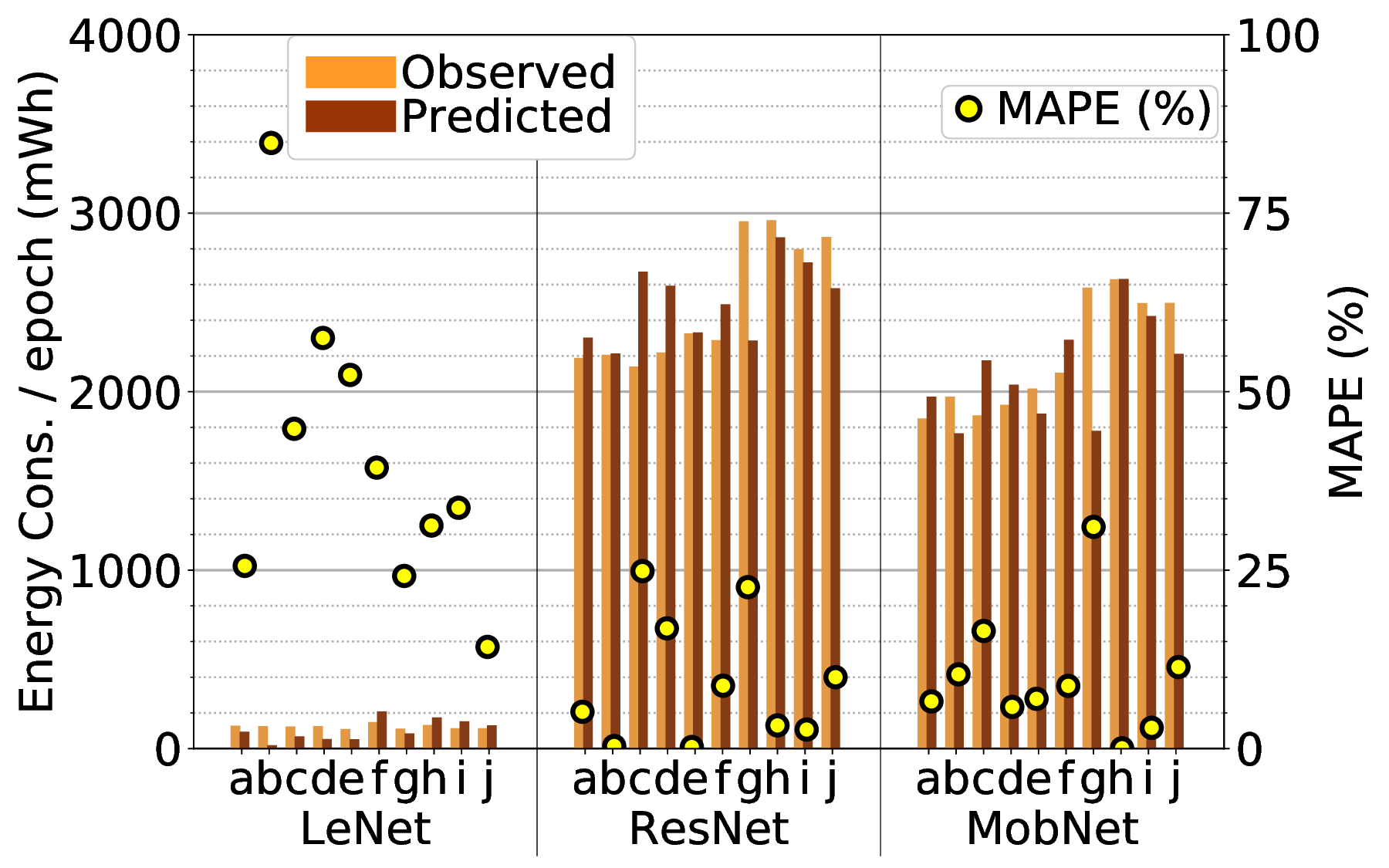}
    \label{fig:predict:energy:mode}
  }
\caption{Predicting end-to-end (E2E) training time and energy use per epoch for power modes.}
\label{fig:predict:e2e-energy}
\end{figure}

\vspace{-1ex}
\subsubsection{Results}
Fig.~\ref{fig:predict:e2e:mode} compares the observed and predicted E2E epoch training times (bars on left Y axis) for the regression model we fit for each of the $3$ DNNs, using the above approach. We evaluate it for the $10$ different power modes, and report the MAPE as a marker on the right Y axis.

Further, we evaluate this approach for a fourth \emph{ab initio} DNN model, VGG11, which was not part of our original evaluation. We fit a regression model for VGG11 using metrics collected for $2$ epochs from the $4$ identified power modes and use it to predict the E2E time for all $10$ power modes.
 
The MAPE\% on the right Y axis of Fig.~\ref{fig:predict:e2e:mode} shows an error of under $10\%$ for 31 out of 40 predictions, and it is within $15\%$ for all but 3 predictions of \lenet. \lenet, the smallest model, exhibits a higher error since our four representative power modes do not have enough variation in their CPU frequency, and \lenet is sensitive to CPU speed due to higher pre-processing and kernel launch times. Interestingly, VGG11 which is a brand-new DNN unknown to our initial modeling shows $<10\%$ error for $9$ out of $10$ power modes. This strengthens our claim that minimal profiling using the $4$ power modes we identify is sufficient to predict the training time for any (non-tiny) DNN.

\vspace{-1ex}
 \subsubsection{Energy}
Similar plots for the energy consumption per epoch are shown in Fig.~\ref{fig:predict:energy:mode}. Here, the prediction is made using the common regression model we provide, which uses the predicted time as an input feature.
The preliminary energy prediction model shows mixed results. While it is able to predict the energy used per epoch for \resnet and \mobilenet within about $25\%$ MAPE, this degrades to about $60\%$ error for \lenet. This is due to the low absolute energy values for \lenet coupled with the higher prediction error in its training time -- which is an input to the energy model. 
Energy prediction for VGG11 also shows poor results, with MAPE values ranging from $21$--$55\%$, and a median of $46\%$ and we do not report it.
The energy modeling requires more investigation.

\section{Discussion and Conclusion}
In this paper, we have conducted a principled study of DNN training on Jetson accelerated edge devices. This exploration is the first of its kind. Our results confirm certain conventional wisdom and back them up with quantifiable metrics. But they also highlight counter-intuitive results which should help rethink system design and tuning for DNN workloads on such platforms.

While the expectation is that caching, pipelining and parallelism of the training workflow should give benefits, this is not always the case. 
An effective drop in end-to-end time is only seen occasionally, where a certain balance between the CPU, GPU and disk speeds, the training data size, and the model size/compute intensity are met.
While one may expect a faster disk to reduce the training time, this too does not always hold. Enabling caching and pipelining can mitigate the effects of a slower disk. 
Mini-batch size should be chosen to maximize GPU parallelism,
but it does not give benefits beyond a point.
Unlike previous studies on inferencing on edge devices, we do not see much variability in training time across device instances, or over time.
The baseload accounts for a large part of training energy, but Wake on LAN can be used to reduce the energy footprint for periodic or on-demand workloads like federated learning.
{Time--energy trade-off} can be exploited for larger DNN models, as seen by the Pareto front, and our training time and energy prediction models offer preliminary but important insights to help exploit the power modes.

We argue that accelerated Jetson edge devices are competitive candidates for DNN training. 
However, effective use of these platforms requires careful tuning and possibly even a redesign of the DNN training platform, which can be guided by our profiling and analysis. 

\vspace{1ex}
\noindent \textbf{Acknowledgments.} We thank the members of the DREAM:Lab including H. Gupta for their help with the paper. We acknowledge the constructive feedback from the reviewers and the shepherd. The first author was supported by a PMRF Fellowship. This work was supported by a DST grant.

\clearpage
\bibliographystyle{IEEEtran}
\bibliography{arxiv}

\input{appendix}
\end{document}

%% file: appendix.tex
\appendix
\clearpage
\section{Appendix}

\begin{figure}[h!]
\centering

  \centering
	\includegraphics[width=0.5\columnwidth]{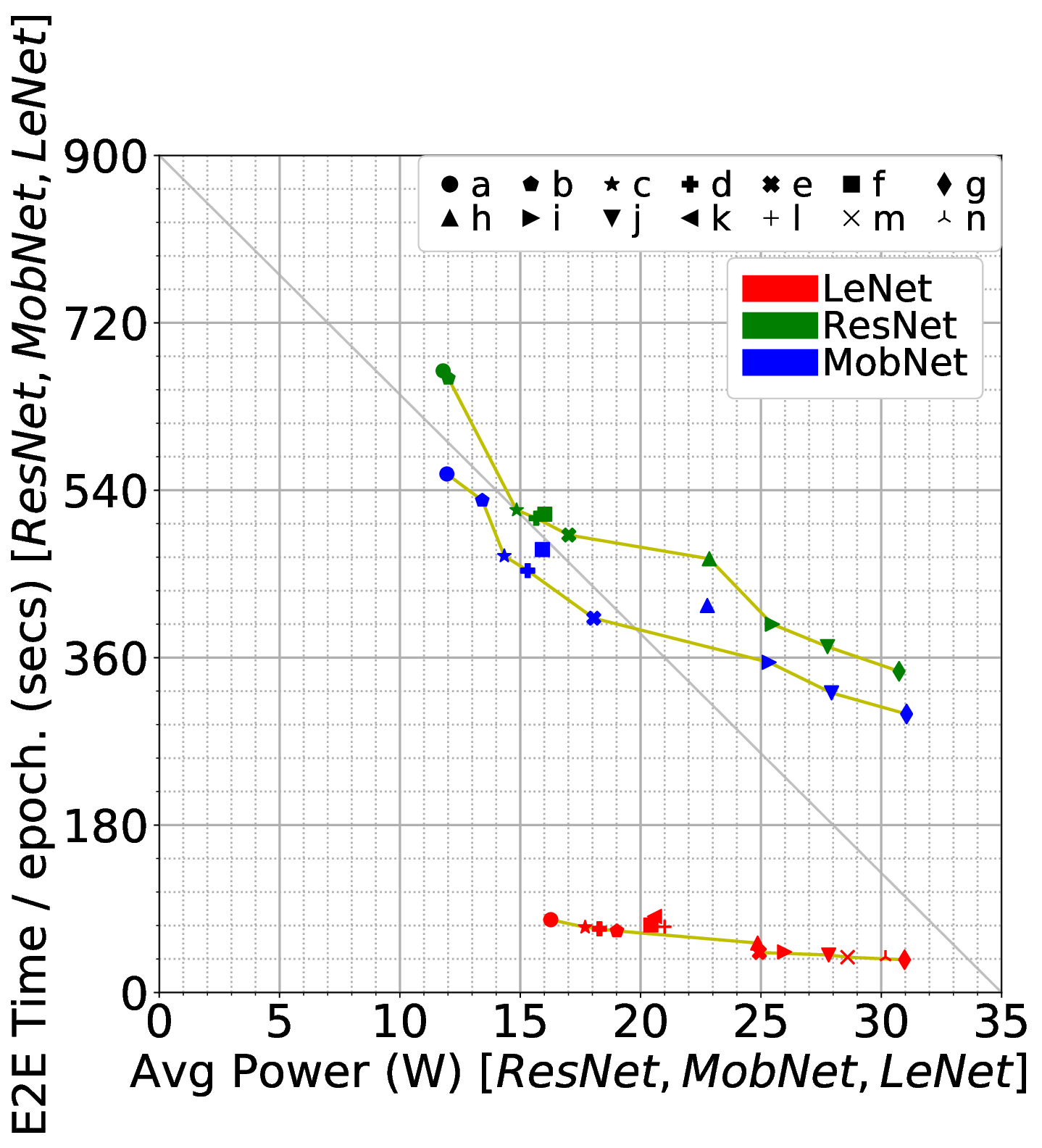}
\caption{Scatter plot of \textit{E2E time vs. Avg power per epoch ($1+$)}, for power modes $a$--$n$ of AGX for \lenet (secondary axes) and \resnet/\mobilenet (primary axes). The yellow lines indicate the \textit{Pareto front} per DNN.}
\label{fig:power:time:scatter}
\end{figure}

\begin{figure}
  \centering
  \includegraphics[width=0.5\columnwidth]{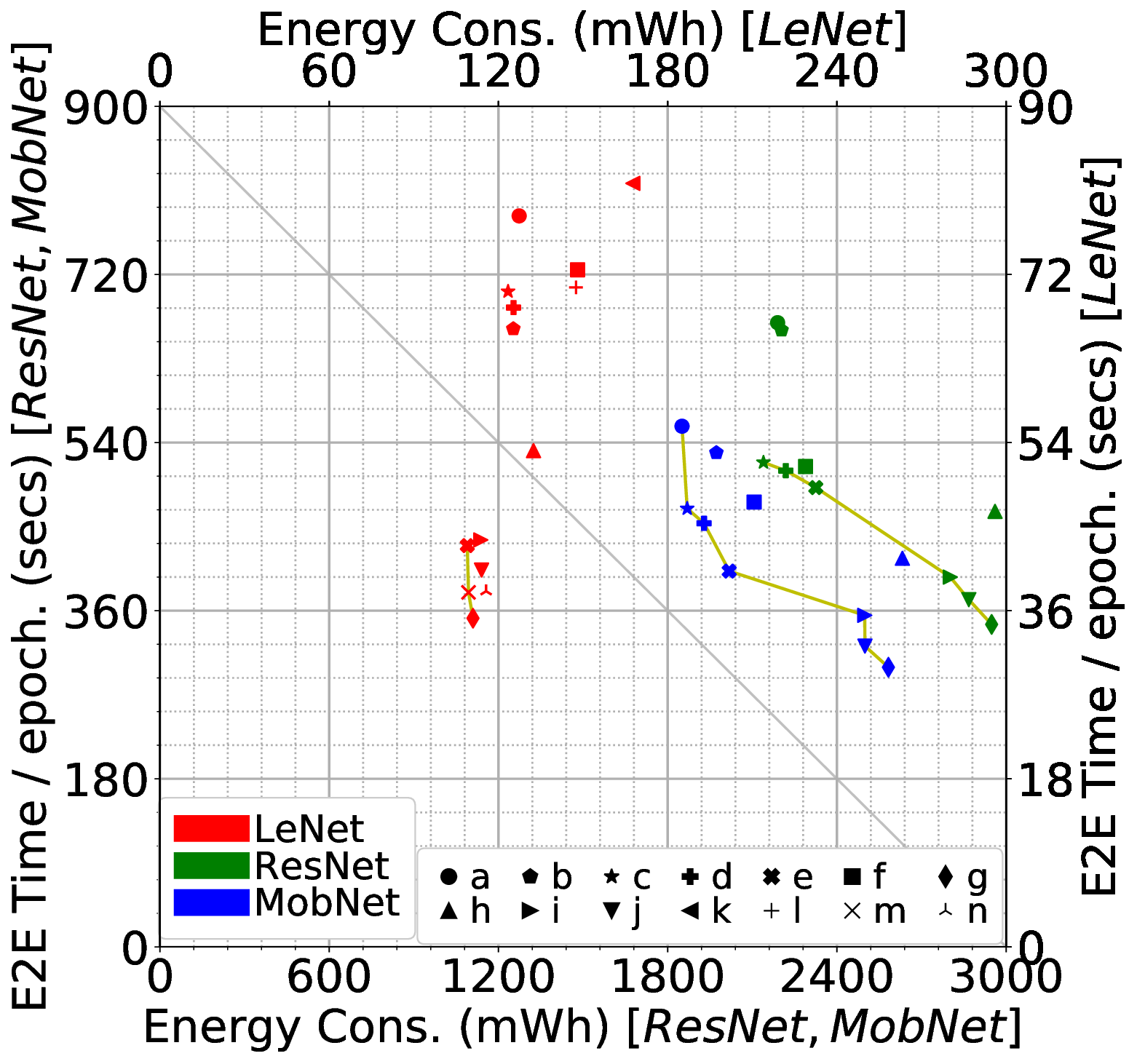}
\caption{Scatter plot of \textit{E2E time vs. Energy consumed per epoch ($1+$)}, for power modes $a$--$n$ of AGX for \lenet (secondary axes) and \resnet/\mobilenet (primary axes). The yellow lines indicate the \textit{Pareto front} per DNN. (Same as Fig.~\ref{fig:power:scatter})}
\label{fig:energy:time:scatter:duplicate}
\end{figure}

In Fig.~\ref{fig:power:time:scatter} we plot the E2E time per Epoch in seconds (Y axis) against the Average Power in Watts (X axis). This complements the plot in Fig.~\ref{fig:power:scatter} (same as Fig.~\ref{fig:energy:time:scatter:duplicate}) which reports Total Energy Consumed in milli-Watt hour (mWh). We see an inverse correlation between epoch time and average.

\begin{figure}
  \centering
	\includegraphics[width=0.5\columnwidth]{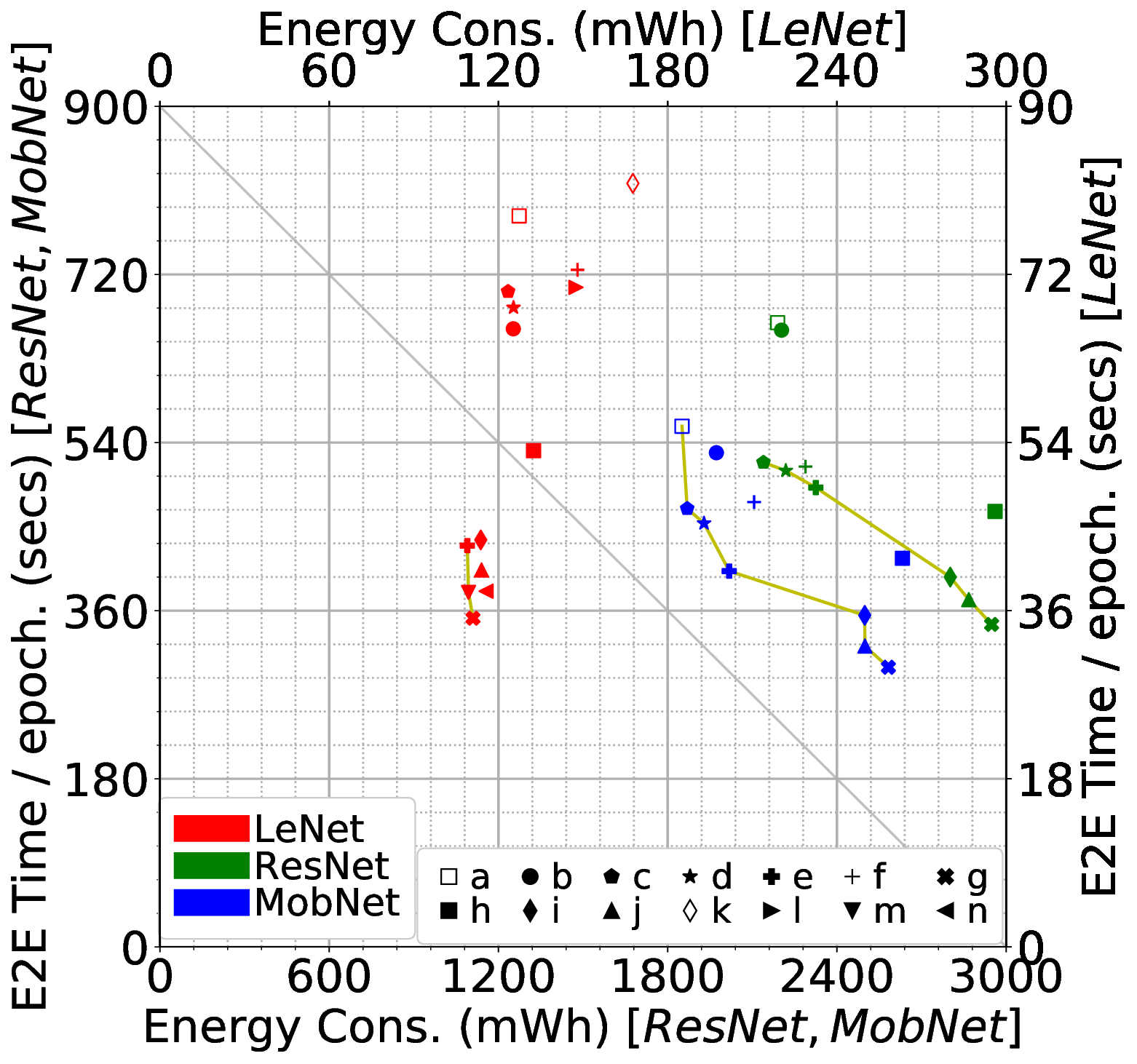}
\caption{Scatter plot and Pareto front of \textit{E2E time vs. Energy consumed per epoch ($1+$)} for AGX for \lenet (secondary axes) and \resnet/\mobilenet (primary axes).\\
Markers are grouped by \textbf{CPU Core Count}.
}
\small
\begin{tabular}{c|L{4.2cm}}
\hline
\# Cores & Marker\\
\hline
\hline 
8 & $\bullet$b\quad $\pentagofill$c\quad \ding{72}d\quad \ding{58}e\quad \ding{54}g\quad $\blacksquare$h\quad $\blacklozenge$i\quad $\blacktriangle$j\quad $\RHD$l\quad $\blacktriangledown$m\quad $\LHD$n \\ 
\hline 
4 & $\square$a\quad $\lozenge$k \\
\hline 
2 & +f\\
\hline
\end{tabular}
\label{fig:energy:time:cores}
\end{figure}

\begin{figure}
  \centering
	\includegraphics[width=0.5\columnwidth]{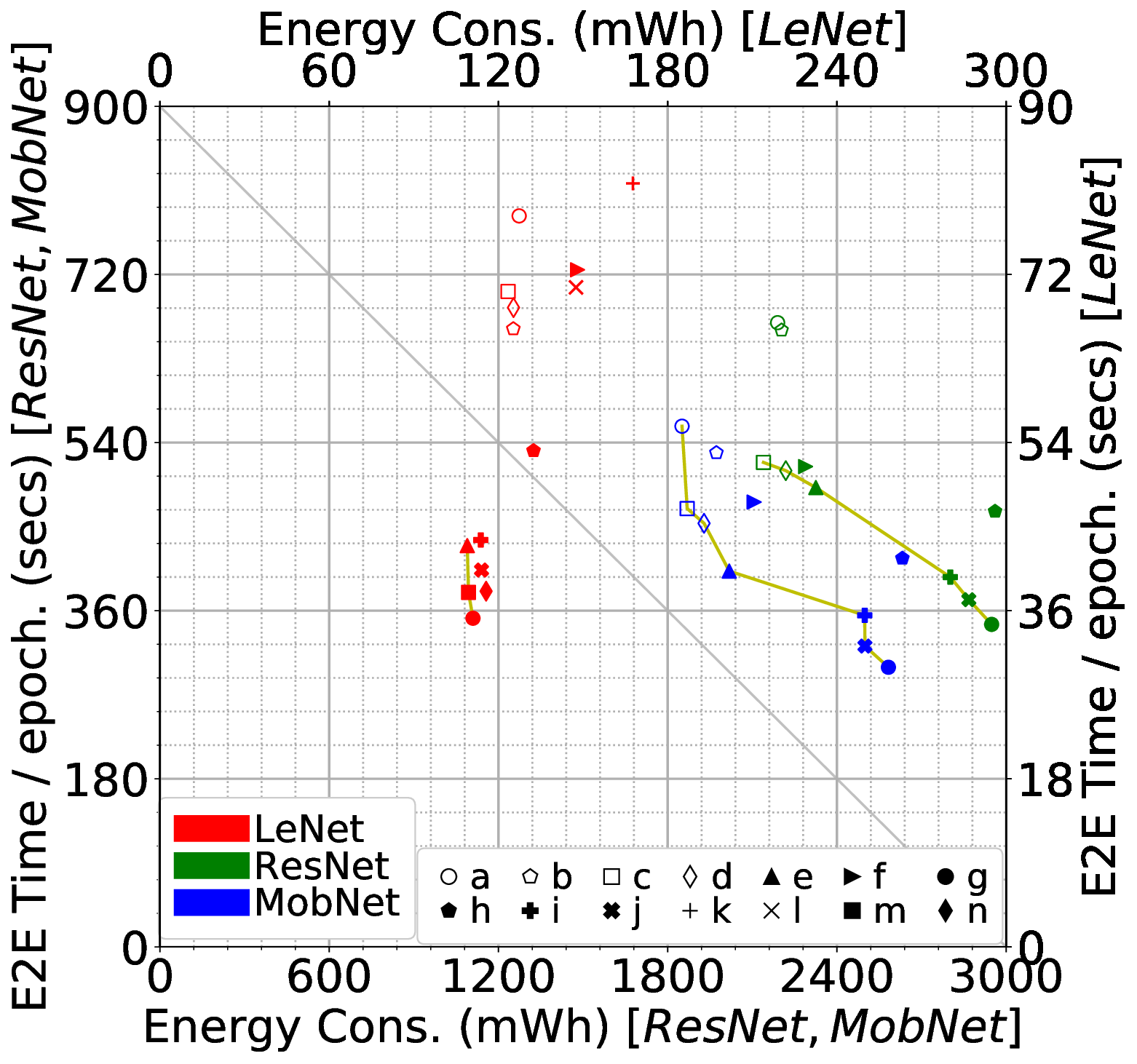}

\caption{Scatter plot and Pareto front of \textit{E2E time vs. Energy consumed per epoch ($1+$)} for AGX for \lenet (secondary axes) and \resnet/\mobilenet (primary axes).\\
Markers are grouped by \textbf{CPU Frequency}.}

\small
\begin{tabular}{C{3cm}|L{3.8cm}} 
\hline
CPU Freq. (MHz) & Marker \\ 
\hline
\hline
2265 & $\bullet$g\quad $\pentagofill$h\quad \ding{58}i\quad \ding{54}j\quad $\blacksquare$m\quad $\blacklozenge$n\\
\hline
2100 & $\blacktriangle$e\quad $\RHD$f\\
\hline
1200 & $\circ$a\quad $\pentagon$b\quad $\square$c\quad $\lozenge$d\\
\hline
1036 & +k\quad $\times$l\\
\hline
\end{tabular}
\label{fig:energy:time:cpuf}
\end{figure}

\begin{figure}
  \centering
	\includegraphics[width=0.45\columnwidth]{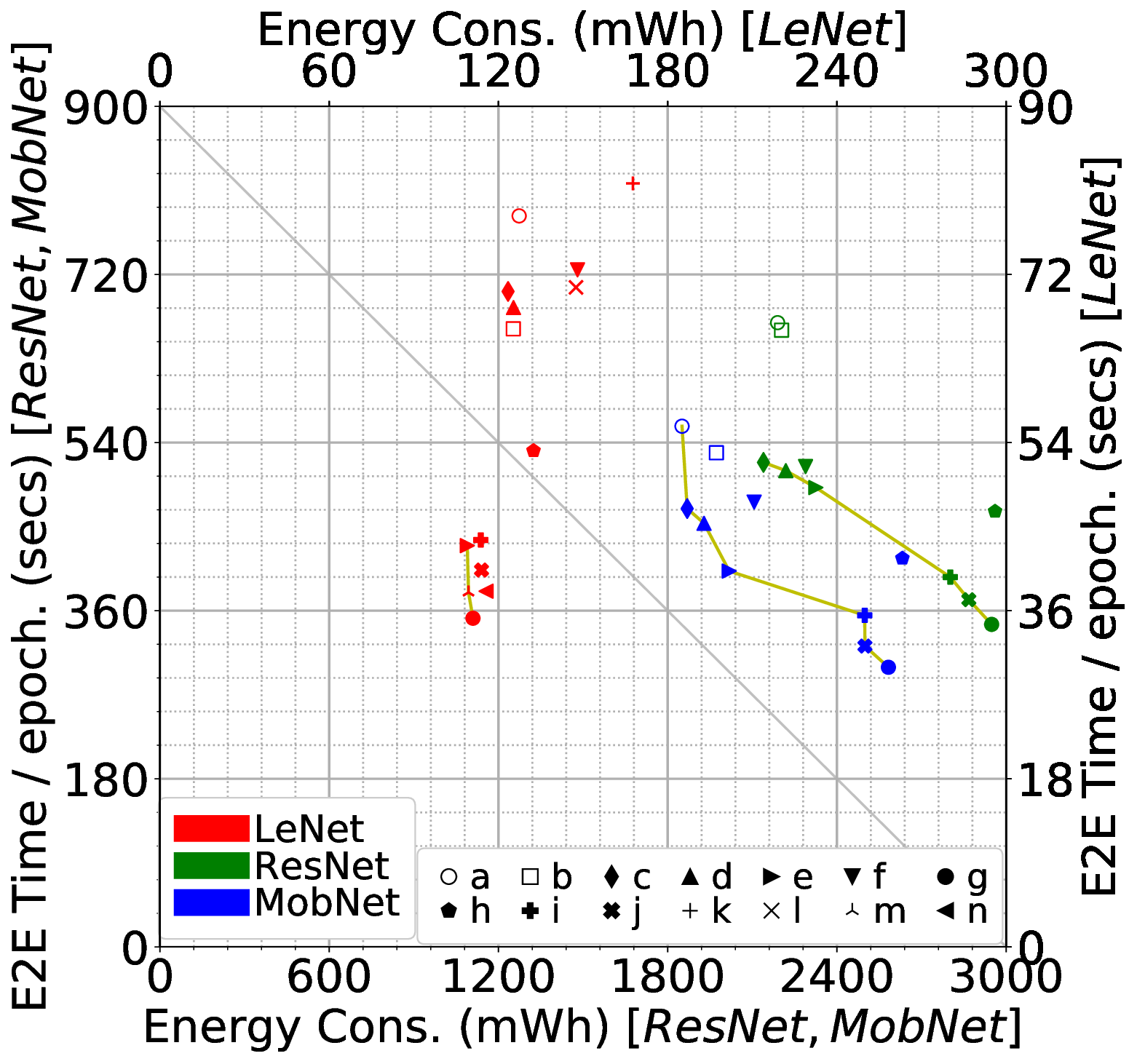}
\caption{Scatter plot and Pareto front of \textit{E2E time vs. Energy consumed per epoch ($1+$)} for AGX for \lenet (secondary axes) and \resnet/\mobilenet (primary axes).\\
Markers are grouped by \textbf{GPU Frequency}.}
\small
\begin{tabular}{C{3cm}|L{3.8cm}} 
\hline
GPU Freq. (MHz) & Marker \\ 
\hline
\hline
1377 & $\bullet$g\quad $\pentagofill$h\quad \ding{58}i\quad \ding{54}j\\
\hline
900 & $\blacklozenge$c\quad $\blacktriangle$d\quad $\RHD$e\quad $\blacktriangledown$f\quad $\LHD$n\\
\hline
670 & $\circ$a\quad $\square$b\\
\hline
420 & +k\quad $\times$l\quad $ \curlywedge$m\\
\hline
\end{tabular}
\label{fig:energy:time:gpuf}
\end{figure}

\begin{figure}
  \centering
	\includegraphics[width=0.45\columnwidth]{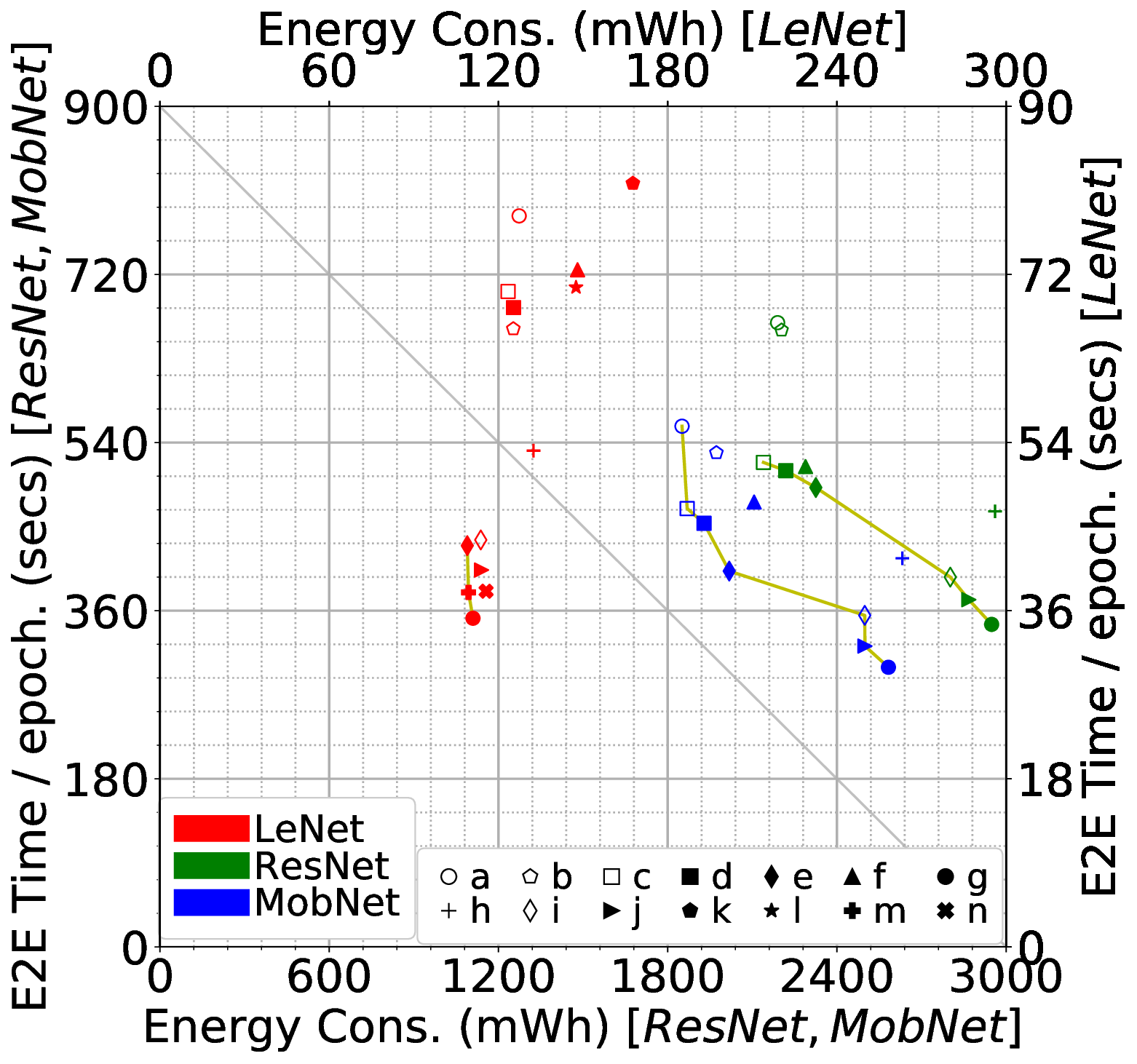}

\caption{Scatter plot and Pareto front of \textit{E2E time vs. Energy consumed per epoch ($1+$)} for AGX for \lenet (secondary axes) and \resnet/\mobilenet (primary axes).\\
Markers are grouped by \textbf{EMC memory Frequency}.}
\small
\begin{tabular}{C{3cm}|L{3.8cm}} 
\hline
EMC Freq. (MHz) & Marker \\ 
\hline
\hline
2133 & $\bullet$g\quad $\pentagofill$k\quad $\bigstar$l\quad \ding{58}m\quad \ding{54}n\\
\hline
1600 & $\blacksquare$d\quad $\blacklozenge$e\quad $\blacktriangle$f\quad $\RHD$j\\
\hline
1333 & $\circ$a\quad $\pentagon$b\quad $\square$c\quad $\lozenge$i\\
\hline
1066 & +h\\
\hline
\end{tabular}
\label{fig:energy:time:memf}
\end{figure}

We reproduce Fig.~\ref{fig:power:scatter} from the main text as Fig.~\ref{fig:energy:time:scatter:duplicate} in the appendix. We further plot $4$ variations of this, each highlighting one of the resources changing (core count, CPU frequency, GPU frequency, EMC memory frequency) and the impact each has on the time--energy trade-off and the Pareto front. For each of these plots, we use a group of related markers to highlight a particular resource value, e.g., solid markers for core count 8, hollow markers for core count 4 and line markers for core count 2 in Fig.~\ref{fig:energy:time:cores} where we focus on the effect of CPU cores. Similarly, Fig.~\ref{fig:energy:time:cpuf} shows the effect of CPU frequency, Fig.~\ref{fig:energy:time:gpuf} the effect of GPU frequency and Fig.~\ref{fig:energy:time:memf} the effect of memory frequency.

\claim{Impact of cores}
The number of cores affects the stall time and kernel launch times. However, this is not a significant component of the E2E time for larger models like \mobilenet and \resnet and only affects \lenet noticeably. For instance, the increase in cores from $f$(+) to $e$(\ding{58}) causes a sharp drop in E2E time and energy for \lenet as seen in Fig.~\ref{fig:energy:time:cores}.

\claim{Impact of CPU frequency}
 The CPU frequency affects the stall time and kernel launch time, and therefore has a larger impact on \lenet's E2E time than the other 2 models. This can be seen from Fig.~\ref{fig:energy:time:cpuf}, where the E2E time values for \lenet corresponding to $2265MHz$ (solid red markers with more than $3$ sides) are much lower than those corresponding to $1200MHz$ (hollow red markers), whereas for \resnet and \mobilenet, the E2E time values of the solid green/blue markers are relatively close to those of the hollow green/blue ones.

\claim{Impact of GPU frequency}
Fig.~\ref{fig:energy:time:gpuf} shows the overall impact that GPU frequency has on the different models. GPU frequency affects the GPU compute time, and therefore the E2E time. This impact is more pronounced for compute-intensive models such as \resnet. As seen in Fig.~\ref{fig:energy:time:gpuf}, for \resnet, there is a significant difference in the E2E time values between the data points with a higher GPU frequency of $900MHz$ (indicated by solid green markers with $3$ or $4$ sides) as compared to those with a GPU frequency of $670MHz$ (hollow green markers). This difference is much lesser for \mobilenet, and even more so for \lenet.

\claim{Impact of memory frequency}
Fig.~\ref{fig:energy:time:memf} shows that memory frequency does have an impact on E2E time and energy, but it is not a dominant factor. The impact of memory frequency can only be seen in modes $g$ to $j$, where all other parameters are kept constant. For instance, the increase in memory frequency in going from $h$(+) to $i$($\lozenge$) causes all $3$ models to see a lower E2E time.

